\documentclass{jfm}
\usepackage{graphicx}
\usepackage{epstopdf, epsfig}
\pdfoutput=1
\usepackage{amssymb}
\usepackage{amsmath}

\usepackage[normalem]{ulem}
\usepackage{color}

\usepackage{tikz}
\DeclareRobustCommand\full  {\tikz[baseline=-0.6ex]\draw[thick] (0,0)--(0.5,0);}
\DeclareRobustCommand\dashed{\tikz[baseline=-0.6ex]\draw[thick,dashed] (0,0)--(0.54,0);}

\usepackage{hyperref}
\usepackage[percent]{overpic}

\usepackage{mylines}
\usepackage{myplots}
\usepackage{layouts}

\newcommand{\lsy}[3]{\mbox{(\kern-.4em\lineSymbolRGB[#3]{#1}{.8pt}{#2}{4pt}\kern-.4em)}}
\newcommand{\sy}[2]{\mbox{(\kern-.25em\SymbolRGB[solid]{#1}{.8pt}{#2}{4pt}\kern-.25em)}}
\newcommand{\lcap}[2]{~\,{\kern-1em\protect\mylcap{#1}{#2}}}

\definecolor{blue}{rgb}{0,0,1}
\definecolor{red}{rgb}{1,0,0}
\definecolor{black}{rgb}{0,0,0}
\definecolor{FCN-POD}{rgb}{0.227,0.454,0.674}
\definecolor{fcn}{rgb}{0.925,0.513,0.207}
\definecolor{epod}{rgb}{0.313,0.607,0.239}

\shorttitle{Wall-bounded turbulence from wall quantities}
\shortauthor{L. Guastoni and others}

\title{Convolutional-network models to predict wall-bounded turbulence from wall quantities}

\author{Luca~Guastoni\aff{1,2}\corresp{\email{guastoni@mech.kth.se}}, 
Alejandro~G\"uemes\aff{3},
Andrea~Ianiro\aff{3},
Stefano~Discetti\aff{3},
Philipp~Schlatter\aff{1,2}, 
Hossein~Azizpour\aff{4,2} \\
\and Ricardo~Vinuesa\aff{1,2}
}

\affiliation{
\aff{1}SimEx/FLOW, Engineering Mechanics, KTH Royal Institute of Technology, \\ Stockholm, Sweden
\aff{2}Swedish e-Science Research Centre (SeRC), Stockholm, Sweden
\aff{3}Aerospace Engineering Research Group, Universidad Carlos III de Madrid, Legan\'es, Spain
\aff{4}Division of Robotics, Perception, and Learning, KTH Royal Institute of Technology, Stockholm, Sweden
}

\begin{document}

\maketitle

\begin{abstract}
Two models based on convolutional neural networks are trained to predict the two-dimensional velocity-fluctuation fields at different wall-normal locations in a turbulent open channel flow, using the wall-shear-stress components and the wall pressure as inputs. The first model is a fully-convolutional neural network (FCN) which directly predicts the fluctuations, while the second one reconstructs the flow fields using a linear combination of orthonormal basis functions, obtained through proper orthogonal decomposition (POD), hence named FCN-POD. Both models are trained using data from two direct numerical simulations (DNS) at friction Reynolds numbers $Re_{\tau} = 180$ and 550. Thanks to their ability to predict the nonlinear interactions in the flow, both models show a better prediction performance than the extended proper orthogonal decomposition (EPOD), which establishes a linear relation between input and output fields. The performance of the various models is compared based on predictions of the instantaneous fluctuation fields, turbulence statistics and power-spectral densities. The FCN exhibits the best predictions closer to the wall, whereas the FCN-POD model provides better predictions at larger wall-normal distances. We also assessed the feasibility of performing transfer learning for the FCN model, using the weights from $Re_{\tau}=180$ to initialize those of the $Re_{\tau}=550$ case. Our results indicate that it is possible to obtain a performance similar to that of the reference model up to $y^{+}=50$, with $50\%$ and $25\%$ of the original training data. These non-intrusive sensing models will play an important role in applications related to closed-loop control of wall-bounded turbulence. 
 
\end{abstract}


\section{Introduction}

The advent of new powerful deep neural networks~\cite[DNNs, see][]{lecun2015deep} has fostered their application in many research areas~\citep{jean_et_al_2016,defauw_et_al_2018,norouzzadeh_et_al_2018,ham_et_al_2019,udrescu,vinuesa_et_al_2020}. Due to their potential applications in flow modelling, identification of turbulence features and flow control, DNNs have recently received extensive attention in the fluid-mechanics research community~\citep{kutz2017deep,jimenez2018machine,duraisamy_et_al,brunton2020machine}. In the case of turbulence modelling, DNNs have been reported to improve the results of Reynolds-averaged Navier--Stokes~\cite[RANS, ][]{ling2016reynolds,wu_et_al} models and large-eddy simulations~\cite[LES, ][]{maulik2019subgrid,lapeyre_et_al,beck_et_al}. There are also a number of on-going efforts towards including the constraints from the Navier--Stokes equations into prediction models through the so-called physics-informed neural networks~\citep{wang2017physics,raissi2019physics}. Furthermore, several artificial-intelligence-based solutions have been proposed to perform optimal control on different types of flows, such as the wake behind one or more cylinders~\citep{rabault2019artificial,raibaudo2020machine}. Other promising applications of machine learning to fluid mechanics include generation of inflow conditions~\citep{fukami_et_al} and extraction of flow patterns~\citep{raissi_et_al}.

DNNs have also found application in temporal prediction of dynamical systems. As an example, \citet{srinivasan2019predictions} compared the capabilities of the multi-layer perceptron (MLP, also known as fully-connected-layer neural network) and several long-short-term memory (LSTM) networks to predict the coefficients of a low-order model for near-wall turbulence~\citep{moehlis_et_al}. While the most relevant flow features are captured by both architectures, the LSTM network outperformed the MLP in terms of ability to predict turbulence statistics and the dynamics of the flow. This work has been extended by \citet{eivazi2020recurrent}, where the LSTM network has been compared with a Koopman-based framework which accounts for non-linearities through external forcing. Although both approaches provide accurate predictions of the dynamical evolution of the system, the latter outperforms the LSTM in terms of time and data required for training. Similar temporal predictions of the near-wall model~\citep{moehlis_et_al} were conducted by \cite{pandey_et_al} using echo state networks (ESN). Moreover, nonlinear autoregressive exogeneous networks (NARXs) have been used by \citet{lozano2020causality} to exploit the relation between the temporal dynamics of the Fourier coefficients of a minimal turbulent channel flow. Their results showed  accurate predictions of the bursting events in the logarithmic layer from buffer-region data. Other related work, in the context of control of the Kuramoto--Sivashinsky (KS) chaotic system, was recently conducted by \cite{bucci_et_al}. Note however that the use of temporal sequences implies a high computational cost to generate well-resolved temporal data. Furthermore, longer sequences require higher memory requirements in order to predict the future behaviour. For these reasons, several neural-network-based models that learn spatial relations have been proposed in the literature. Convolutional neural networks (CNNs) have become increasingly popular during the last years due to the hierarchical structure of their input~\citep{fukushima1980neocognitron,fukushima1988neocognitron, lecun1989backpropagation,Lecun98gradient-basedlearning}. For instance, \citet{fukami2019super,fukami2020machine} have shown that turbulent flow fields can be reconstructed from extremely coarse data with remarkable success.
CNNs have also been used to investigate the dynamical features of the flow without \textit{a-priori} knowledge, as shown by \citet{jagodinski2020uncovering}.

Neural networks are mathematical models based on data-driven training, and as such they have been compared and used together with other data-driven methods. For instance, the relationship between proper orthogonal decomposition~\cite[POD, see][]{lumley1967structure} and the MLP is well documented in the literature~\citep{bourlard1988auto,baldi1989neural}. These works showed that a MLP with a single hidden layer is equivalent to POD if a linear activation function is used. \citet{milano2002neural} compared the results of POD-based neural networks with linear and non-linear functions for the prediction of near-wall velocity fields, showing that nonlinear POD has significantly better predictive capabilities. More recently, the emergence of autoencoder architectures has motivated a renewed interest in the application of neural networks for dimensionality reduction. \citet{hinton2006reducing} proposed the use of deep autoencoders to obtain a low-order representation of high-dimensional data, showing that this approach is able to retain more information than POD. It is interesting to note that this work avoids the inherent difficulty of optimizing weights in deep autoencoders by training each layer with a Restricted Boltzmann Machine. \citet{murata2020nonlinear} used an autoencoder with convolutional layers to obtain a low-order representation of the flow around a cylinder. Their results suggest that CNN autoencoders with linear activation functions reproduce the same dimensionality reduction as POD, while the use of nonlinear activation functions improves the reconstruction performance. On a related note, flow reconstruction based on shallow neural networks was studied by \cite{erichson_et_al} in several fluid-mechanics examples.

In this work, we assess the potential of DNNs for {\it non-intrusive sensing}, to be used for closed-loop control applications. In this type of control, the actuation is applied with the aim of suppressing the effect of certain structures in the flow~\citep{choi_et_al}. In order to effectively perform closed-loop control it is necessary to monitor the instantaneous state of the flow so as to devise the best way to affect it, but this type of measurement is extremely challenging, particularly at very high Reynolds numbers where the near-wall structures become progressively smaller. On the other hand, it is more feasible to perform non-intrusive sensing, {\it i.e.} to accurately measure time-resolved quantities at the wall, such as the wall-shear stress or the pressure, and then correlate these measurements with the flow farther away. In a seminal study over 20 years ago, \cite{lee_et_al_cnn} used a CNN to predict, based on the wall-shear-stress components, the wall actuation that would lead to higher drag reduction. More recently, \citet{guastoni} used the two wall-shear-stress components to predict the instantaneous streamwise flow fields at several wall-normal positions using convolutional networks. Their results show that these neural networks provide better predictions than linear methods (see below) in terms of instantaneous predictions and second-order statistics. The same wall information was used by \citet{kim2020prediction} to predict the instantaneous wall-normal heat flux  with satisfactory results. Moreover, in the work by \citet{Guemes2019sensing} the information of the most-energetic scales was encoded into a POD basis, and a CNN was used to predict that information at different wall-normal locations from streamwise wall-shear-stress measurements. Their results demonstrated that CNNs can significantly outperform linear methods in the prediction of POD time coefficients for low-order reconstruction of the velocity fields.

A drawback of DNNs is the fact that they require training and test data to be taken from the same distribution, {\it i.e.} for the same flow and at the same Reynolds number in our case. However, in  a real-world application the flow conditions will be continuously varying and/or it might be unfeasible to perform a full training at exactly the same conditions. It may be possible, however, to exploit initial training at a certain flow condition and transfer this knowledge to another one. Such knowledge transfer could reduce significantly the amount of data needed for training and improve the network applicability for industrial applications. Transfer learning~\citep{pan2009survey} is the suitable learning framework to address this issue, and it is discussed in detail below.

Before the appearance of DNNs, flow-field prediction was performed mainly through linear methods. Among them, the linear stochastic estimation (LSE) introduced by \citet{adrian88} stands out. Recently \citet{suzuki_hasegawa_2017} and \citet{encinar} have used LSE to reconstruct the velocity field on a wall-parallel plane in a turbulent channel flow employing wall measurements. The latter study showed that LSE can only reconstruct the large wall-attached eddies in the outer part of the logarithmic region. An extension of the LSE method in the spectral domain~\citep{tinney2006spectral} was shown to be more suitable for noisy predictions in turbulent flows. More recently \citet{baars2014proper} proposed a POD-based method for improving the spectral-LSE approach. \citet{boree2003extended} reported the possibility of projecting a synchronized field on the POD temporal modes of another quantity; this method is known as extended POD (EPOD). The correlation matrix between the temporal POD coefficients of two given quantities can be used to predict one based on the other one. The work of \citet{boree2003extended} proved EPOD to be equivalent to LSE when all modes are retained in the reconstruction. EPOD has been used to provide predictions in turbulent jets~\citep{tinney2008low}, channel flows~\citep{discetti2018estimation}, pipe flows~\citep{discetti2019characterization} and wall-mounted obstacles~\citep{bourgeois2013generalized,hosseini_martinuzzi_noack_2016} using remote probes. On the other hand, \citet{sasaki_vinuesa_cavalieri_schlatter_henningson_2019} recently assessed the capabilities of both linear and non-linear transfer functions with single and multiple inputs to provide turbulent-flow predictions. They documented a significant improvement in the predictions when the transfer functions were designed to account for nonlinear interactions between the inputs and the flow field. The improved prediction capabilities of nonlinear methods over linear ones were also reported by \cite{mokhasi_et_al} and \cite{nair_goza}.

The methods proposed by \citet{guastoni} and \citet{Guemes2019sensing}, henceforth referred to as fully-convolutional network (FCN) and FCN-POD respectively, are extended in the present study. Both models are able to provide a nonlinear characterization of the relation between wall features and the flow on wall-parallel planes. The purpose of this work is to provide a detailed comparison of the two aforementioned nonlinear methods regarding their capabilities to predict turbulent flow fields from wall information. Their improvement over linear methods is measured using EPOD. Furthermore, transfer learning was applied to the FCN approach  with the purpose of evaluating to what extent a network trained at one Reynolds number can be used at a different one.
 
The remainder of this article is organised as follows. Section \ref{datasets} describes the numerical databases used for training and testing the neural networks and $\S$\ref{cnn} provides a brief introduction to convolutional neural networks. The FCN and FCN-POD methods are presented in $\S$\ref{ss:FCN} and $\S$\ref{ss:FCN-POD} respectively, while EPOD is discussed in $\S$\ref{ss:epod}. Results from the considered prediction methods are presented and compared in $\S$\ref{ss:results}, including instantaneous fields in $\S$\ref{ss:inst}, second-order statistics in $\S$\ref{ss:stat}, and spectra in $\S$\ref{ss:spec}. Furthermore, an assessment of transfer learning between different Reynolds numbers is presented in $\S$\ref{ss:tl}. Finally, the main conclusions of the work are presented in $\S$\ref{ss:conclu}. An Appendix is provided containing additional information regarding the predicted instantaneous flow fields.

\section{Methodology}
\subsection{Datasets}\label{datasets}
All the DNN variants in this study have been trained using the data generated from direct numerical simulations (DNS) of a turbulent open-channel flow. Periodic boundary conditions are imposed in the \textit{x-} and \textit{z-}directions (which are the streamwise and spanwise coordinates, respectively), and a no-slip condition is applied at the lower boundary ($y=0$, where $y$ is the wall-normal coordinate). Differently from a standard channel-flow simulation, a symmetry condition is imposed at the upper boundary. In standard channel flows, the wall-attached coherent structures may extend beyond the channel centerline, thus affecting the other wall~\citep{lozano_2012}.  
On the other hand, in open-channel flows there is no upper wall. This makes the simulation more suitable to investigate to which extent the neural networks are able to learn the dynamics of near-wall turbulence, since the interaction of the large scales with both walls is not present.

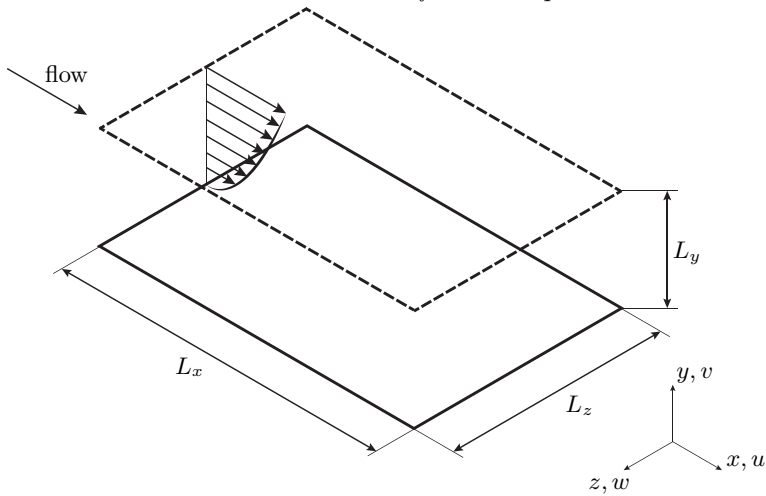
\begin{figure}
\begin{center}
\begin{overpic}[width=0.75\textwidth]{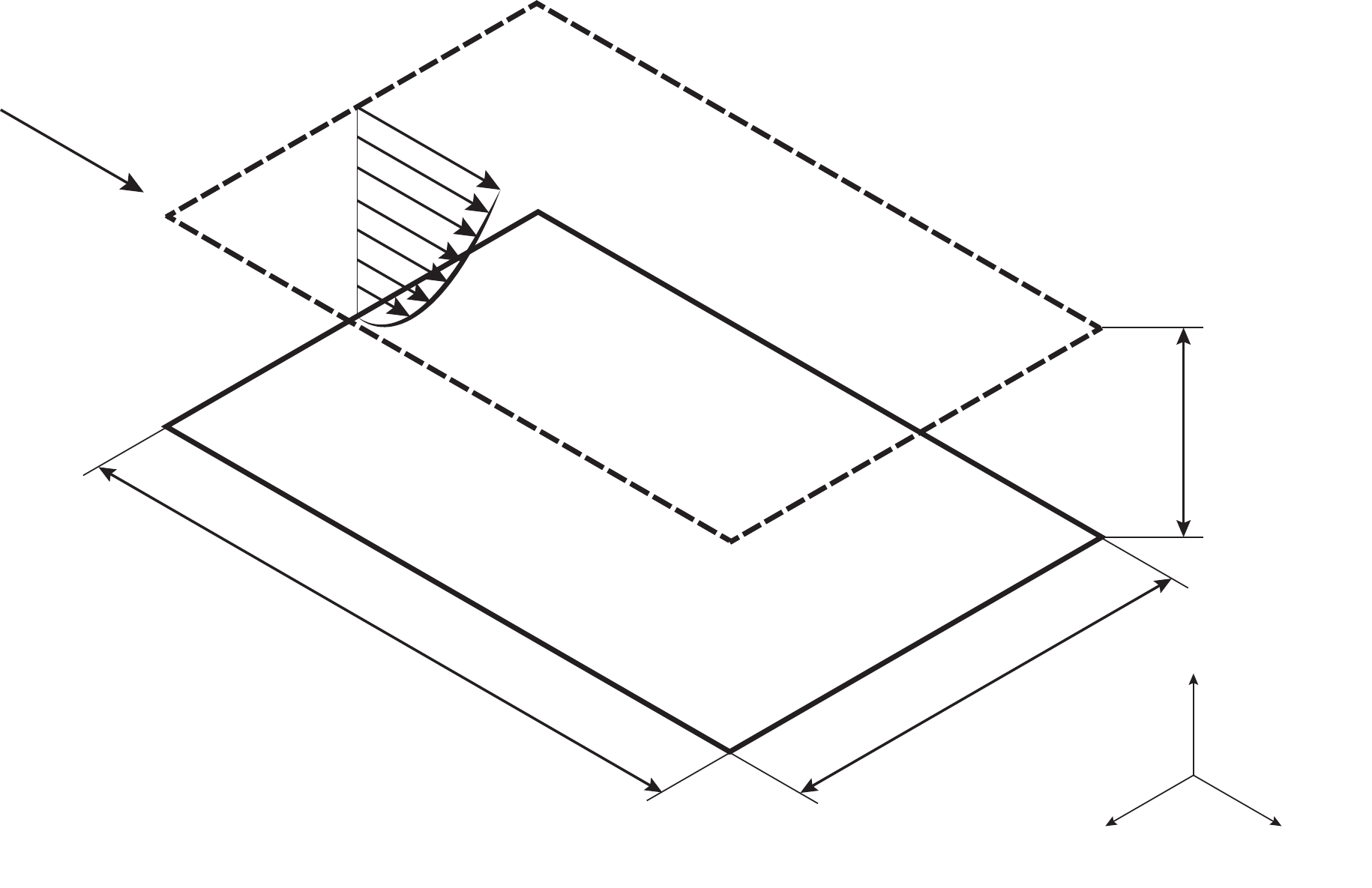}
 \put (5,55) {flow}
 \put (22,17) {$L_x$}
 \put (73,12) {$L_z$}
 \put (87,32) {$L_y$}
 \put (87.5,16.5) {$y,v$}
 \put (94,5) {$x,u$}
 \put (76,2) {$z,w$}
\end{overpic}
\end{center}
\caption{\label{channel}Computational domain and frame of reference for the DNS of the turbulent open channel considered in this study.}
\end{figure}

The simulation is performed using the pseudo-spectral code SIMSON~\citep{chevalier} with constant mass flow rate, in a domain $\Omega = L_x \times L_y \times L_z = 4\pi h \times h \times 2\pi h$ (where $h$ is the channel height), as shown in figure~\ref{channel}. Two friction Reynolds numbers $Re_{\tau}$ (based on $h$ and the friction velocity $u_{\tau}=\sqrt{\tau_w/\rho}$, where $\tau_w$ is the wall-shear stress and $\rho$ is the fluid density) are considered, as summarized in table~\ref{tab:dns}. The flow field is represented with $N_y$ Chebyshev modes in the wall-normal direction and with $N_x$ and $N_z$ Fourier modes in the streamwise and spanwise directions, respectively. 
The instantaneous fields are obtained at constant time intervals from the time-advancing scheme, which is a second-order Crank--Nicholson scheme for the linear terms and a third-order Runge--Kutta method for the nonlinear terms. Dealiasing using a standard 3/2 rule is employed in the wall-parallel Fourier directions. The velocity fields to be used as ground truth for training and testing are sampled at the following inner-scaled wall-normal coordinates: $y^{+}=15,~30,~50$ and $100$. Note that `+' denotes viscous scaling, {\it i.e.} in terms of the friction velocity $u_{\tau}$ or the viscous length $\ell^{*}=\nu / u_{\tau}$ (where $\nu$ is the fluid kinematic viscosity). A dataset is defined as a collection of samples, each consisting of the shear-stress and pressure fields at the wall as inputs, along with the corresponding velocity fields at the target wall-normal locations as outputs. The training/validation dataset at $Re_{\tau} = 180$ consists of 50,400 instantaneous fields, with a sampling period of $\Delta t^+_{s} = 5.08$. The sampling period at $Re_{\tau} = 550$ is set to $\Delta t^+_{s} = 1.49$ and the training/validation dataset includes 19,920 fields. In both cases, the dataset is split into training and validation sets, with a ratio of 4:1. As shown in table~\ref{tab:dns}, the number of Fourier modes in the wall-parallel directions is higher at $Re_{\tau}=550$, even if the the resolution of the individual fields is the same in viscous units. Since the domain much larger when scaled in inner units, a higher number of flow features is sampled per snapshot in the high-$Re_{\tau}$ case, thus partially compensating the lower number of fields.

The predictions used to assess the performance of the trained models were obtained from additional test datasets. This is done both at $Re_{\tau} = 180$ and $Re_{\tau} = 550$, and it is necessary to ensure that the training and test datasets are completely uncorrelated, both in space and time.  Test samples were taken from simulations initialized with a random seed, different from that of the training-data simulation. The size of the test dataset (more than 3,000 flieds for both $Re_{\tau}$) is sufficient to achieve convergence of the turbulence statistics from the predicted flow, and then these are compared  with the reference values from the DNS.

\begin{table}
\begin{center}
    \begin{tabular}{l l*{6}{c}}
        Name   & $Re_{\tau}$ & $N_x \times N_z \times N_y$           & $\#$Train.+Val. fields & $\Delta t^+_{s,\mathrm{train}}$ & $\#$Test fields & $\Delta t^+_{s,\mathrm{test}}$ \\[0.3cm]
        
        DNS180 & 180         & $192 \times 192 \times 65\phantom{0}$ &  50,400          & 5.08                            & 3,125       & 1.69                           \\
        DNS550 & 550         & $512 \times 512 \times 193$           & 19,920          & 1.49                            & 3,320       & 1.49                           \\
        
    \end{tabular}
    \caption{Description of the DNS datasets used for computing the EPOD and training/testing the CNN-based models.}
    \label{tab:dns}
\end{center}
\end{table}

\subsection{Convolutional neural networks}\label{cnn}

In this study we consider the instantaneous two-dimensional fields of the streamwise and spanwise wall-shear-stress components and of the wall pressure as inputs for our models. The presence of coherent features in the input fields motivates the use of convolutional layers in our neural-network models to process the information. In these layers, a convolution in two dimensions is performed, and it is defined as:
\begin{equation}
    F_{i,j} = \sum_m \sum_n I_{i-m,j-n}K_{m,n},
\end{equation}
where $\mathbf{I} \in \mathbb{R}^{d_1 \times d_2}$ is the input, $\mathbf{K} \in \mathbb{R}^{k_1 \times k_2}$ is the so-called \textit{kernel} (or \textit{filter}) containing the learnable parameters of the layer, and the transformed output $\mathbf{F}$ is the \textit{feature map}. Multiple feature maps can be stacked and sequentially fed into another convolutional layer as input. This allows the next layer to combine the features individually identified in each feature map, enabling the prediction of larger and more complex features for progressively deeper convolutional networks. Since $k_i \ll d_i\ \forall i$, the use of kernels greatly reduces the number of parameters that need to be learned during training (in comparison with fully-connected MLP networks).

A DNN that features this kind of layers is known as \textit{convolutional neural network}~\cite[CNN, see][]{Lecun98gradient-basedlearning}. In this work we consider two different architectures to predict the instantaneous velocity fields at different wall-normal locations based on the same input fields. In one case, the instantaneous two-dimensional velocity fluctuations are directly predicted from the input fields by using a fully-convolutional neural network (FCN). This network is similar, but conceptually different from CNNs, which typically have several convolutional layers followed by one or more fully-connected layers (which are the building blocks of MLP networks). In these networks the localized information processed by the individual convolutional kernels is combined to obtain a global prediction, whereas in FCNs only convolutional layers are present and the network architecture is based on the assumption that the relation between input and output variables is spatially localized. The input region from which a single point of the output can draw information is called \textit{receptive field} and it can be computed based on the network architecture, as described by~\citet{dumoulin2016guide}. Additional details regarding this architecture are provided in $\S$\ref{ss:FCN}. The second approach is a development of the one used by \citet{Guemes2019sensing}, and it is based on the following steps: first, the fluctuation fields are projected on an orthonormal basis using proper orthogonal decomposition (POD)~\citep{lumley1967structure}, so that the spatial and temporal dynamics are separated. Then, the neural network reconstructs the velocity fluctuation field by predicting the coefficients that determine the temporal dynamics. Here we also employ a fully-convolutional network, which will be referred to as FCN-POD, and its architecture is described in $\S$\ref{ss:FCN-POD}.

\subsection{Fully-convolutional neural-network predictions}\label{ss:FCN}
FCNs are commonly used in applications where the input and output domains have structural similarities. \textit{Image segmentation}~\citep{long2015fully} is one such case, since the output has the same spatial dimension as the input, as in our predictions of two-dimensional flow fields. The inputs of the network are the wall-shear-stress components in the streamwise and spanwise directions, as well as the pressure at the wall. Each of the inputs of the network is normalized using the respective mean and standard deviation, as computed from the training/validation set. The predictions are performed using the same mean and standard deviation values on the test dataset inputs. The outputs are the instantaneous velocity fluctuations, denoted as $u$, $v$ and $w$ (corresponding to the streamwise, wall-normal and spanwise velocities, respectively), at a given distance from the wall. Note that the predictions are carried out at the same time as the input fields. In our previous work~\citep{guastoni}, a similar FCN was used to predict the streamwise component of instantaneous flow fields. In the present study the predictions are extended to the wall-normal and spanwise components, however the back-propagation algorithm that is used to train the networks works best when the error in the prediction of three outputs (\textit{i.e.} the three velocity components) has a similar magnitude for all of them. Thus, the fluctuations are scaled as follows: 
\begin{equation}
    \widehat{u} = u, \quad \widehat{v} = v \frac{u_\mathrm{RMS}}{v_\mathrm{RMS}}, \quad \widehat{w} = w \frac{u_\mathrm{RMS}}{w_\mathrm{RMS}},
    \label{scaling}
\end{equation}
where RMS refers to root-mean-squared quantities. During inference (\textit{i.e.}\ when the predictions are computed from the inputs in the test dataset), the outputs of the network are re-scaled back to their original magnitude. The network is trained to minimize the following loss function:
\begin{equation} \label{eq:loss}
\mathcal{L}_{{\rm FCN}}(\boldsymbol{\widehat{u}}_\mathrm{FCN};\boldsymbol{\widehat{u}}_\mathrm{DNS})=\frac{1}{N_x N_z} \sum_{i=1}^{N_x} \sum_{j=1}^{N_z} \left | \boldsymbol{\widehat{u}}_\mathrm{FCN}(i,j) - \boldsymbol{\widehat{u}}_\mathrm{DNS}(i,j)\right |^{2},
\end{equation}
which is the mean-squared error (MSE) between the instantaneous prediction $\boldsymbol{\widehat{u}}_\mathrm{FCN}$ and the true velocity fluctuations $\boldsymbol{\widehat{u}}_\mathrm{DNS}$, as computed by the DNS and scaled in the same way as the network outputs.

The chosen inputs and outputs allow the FCN to learn only the spatial relation between the quantities at the wall and the fluctuations farther away from it. Note that it would also be possible to consider predictions in time, and in that case convolutional neural networks could be used~\citep{oord2016wavenet} treating time as another spatial coordinate, or it would be possible to use recurrent networks, specifically designed to learn temporal sequences as we have recently shown with long-short term memory (LSTM) networks~\citep{srinivasan2019predictions,guastoni_tsfp}. In both cases, the need of multiple samples in time makes the model less flexible than one that relies only on spatial correlations, both during training and testing. These models usually assume a constant sampling time for the data sequence, which might be difficult to enforce, for example if the fields are taken from a numerical simulation with adaptive time step. 
During inference, models that work with sequences would require input fields at different times to perform the prediction. On the other hand, a single input sample is sufficient for the FCN to predict the output.

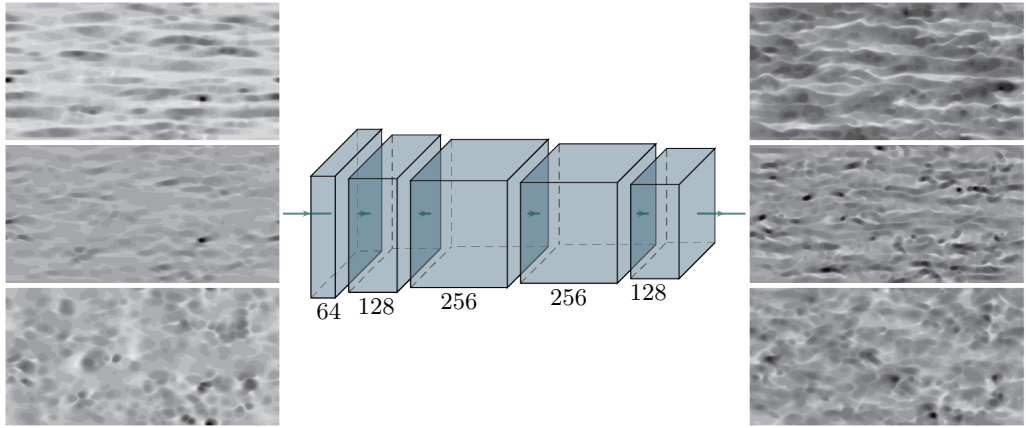
\begin{figure}
\begin{center}
\begin{overpic}[width=\textwidth]{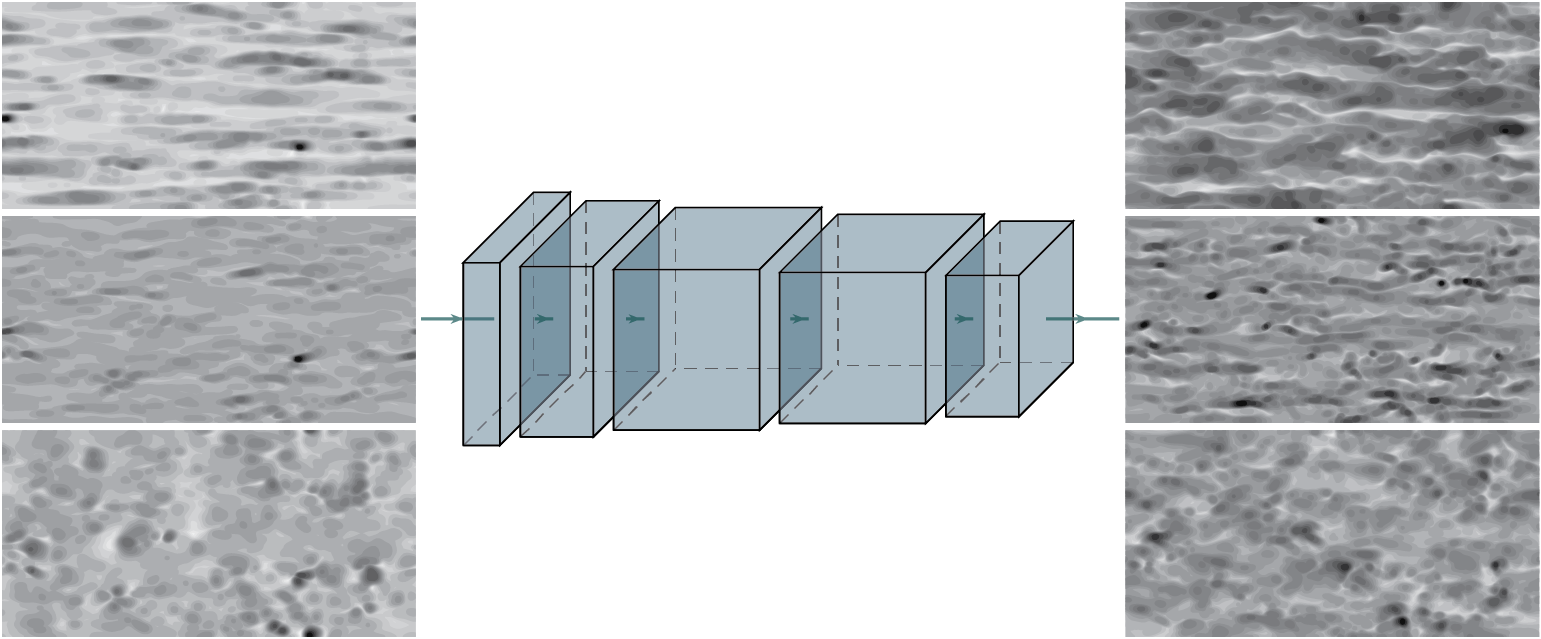}
 \put (30.6,10.5) {$64$}
 \put (34.65,11) {$128$}
 \put (42.75,11.5) {$256$}
 \put (53.5,11.9) {$256$}
 \put (61.3,12.4) {$128$}
\end{overpic}
\end{center}
\caption{\label{fig:net} Schematic representation of the considered FCN architecture. The input fields are on the left (from top to bottom: streamwise wall-shear stress, spanwise wall-shear stress and wall pressure) and the outputs are on the right (from top to bottom: streamwise, wall-normal and spanwise fluctuations). The numbers indicate the number of kernels applied to each of the layers. The kernels (not represented in the figure) have size $(5\times 5)$ in the first convolutional layer, and $(3\times 3)$ in the subsequent layers. A darker colour corresponds to a higher value of the represented quantity.}
\end{figure}
Input and output fields are obtained from a simulation with periodic boundary conditions in the streamwise and spanwise directions. Such constraints could be added to the loss function, however this would imply that periodicity would only be satisfied in a least-square sense. In our implementation we are able to enforce periodicity in both wall-parallel directions by leveraging the fact that the convolutional-network output is deterministic and influenced only by the local information in the receptive field. In other words, if the network receives a certain local input, the local output value will always be the same, regardless of the local position within the input field. In order to have the same values on both edges of the domain, the inputs fields are padded in the periodic directions, {\it i.e.}, they are extended on both ends, using the values from the other side of the fields. 

The FCN architecture is shown in figure~\ref{fig:net}. Each convolution operation (except for the last one) is followed by batch normalization~\citep{ioffe} and a rectified-linear-unit~\cite[ReLU, see][]{nair2010rectified} activation function. The receptive field for this architecture is $15 \times 15$ points, hence 16 points are added to each field in both streamwise and spanwise directions. Note that this padding leads to a network output that is slightly larger than the velocity fields from the DNS, and therefore the network output is cropped to match the size of the reference flow fields. 
The padding involves a computational overhead due to the increased size of the input fields, however it is important to highlight that the padding is architecture-dependent and not input-dependent, meaning that the input is $\approx 17\%$ bigger with a $192 \times 192$ field resolution (at $Re_{\tau} = 180$), but only $\approx 6\%$ bigger when the fields have a size of $512 \times 512$ (at $Re_{\tau} = 550$). Moreover, the FCN was trained using the Adam~\citep{kingmaba} optimization algorithm for 50 epochs, with a scheduled exponential learning-rate decay. We used the optimizer hyperparameters suggested by \cite{kingmaba}. The total number of trainable parameters in the FCN is 1,264,131.

\subsection{POD-based predictions with convolutional neural networks}\label{ss:FCN-POD}

The methodology proposed by \citet{Guemes2019sensing} employs a field of streamwise wall-shear stress to reconstruct the flow field at a certain wall-normal distance as a linear combination of orthogonal modes $\boldsymbol{\phi}_i(\boldsymbol{x})$:
\begin{equation}
    \boldsymbol{u}(\boldsymbol{x},t) \approx \sum^{N_m}_{i=1}a_i(t)\sigma_i\boldsymbol{\phi}_i(\boldsymbol{x}),
    \label{eq:decom}
\end{equation}
where $N_m$ is the total number of POD modes, $a_i(t)$ is the temporal POD coefficient corresponding to mode $i$, and $\sigma_i$ is its corresponding root-squared energy contribution. While the orthogonal modes were estimated from a POD of the training dataset, the network was trained and then employed to predict the mode temporal coefficients for each snapshot. This approach was assumed to be especially advantageous since it allows to filter out the noise content represented by small and uncorrelated scales, thus taking advantage of the energy optimality of POD modes.  

\begin{figure}
    \centerline{\includegraphics[width=384pt]{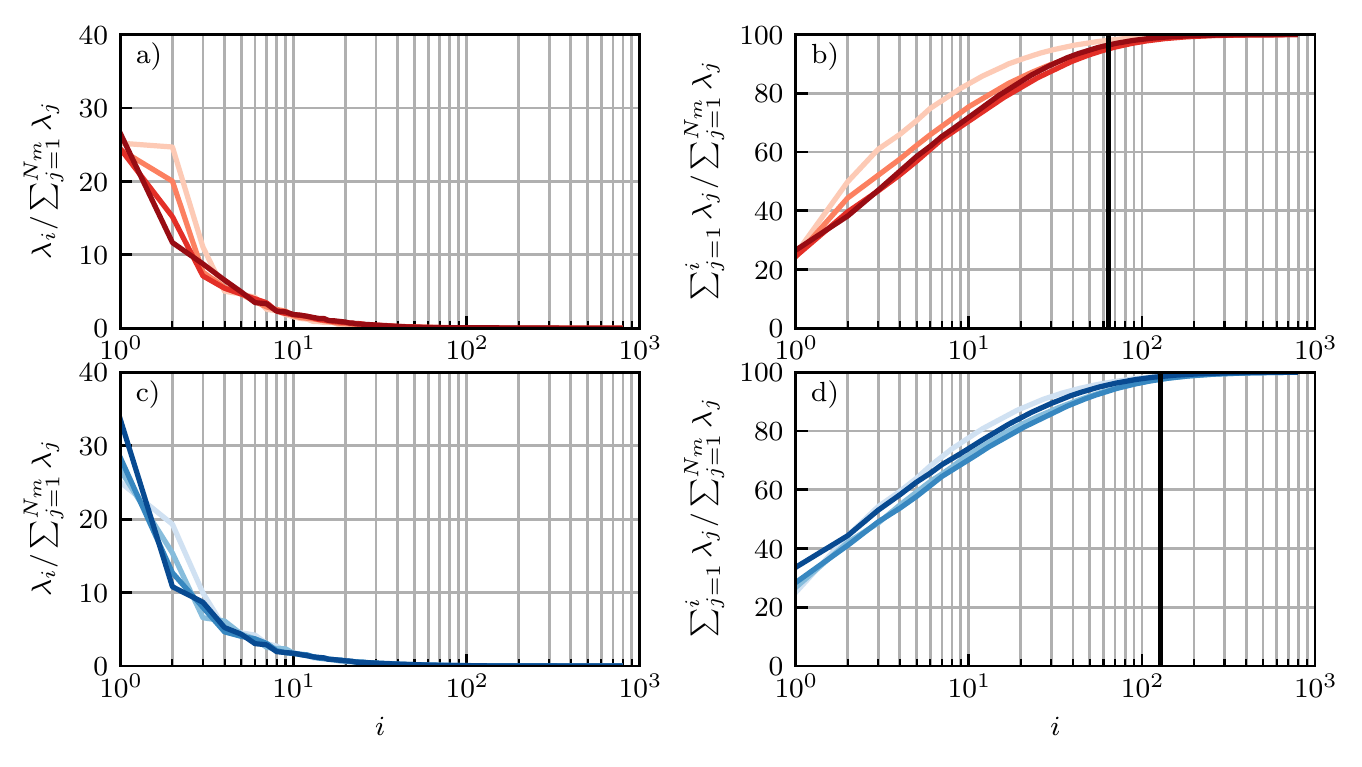}}
    \caption{Distribution of a), c) POD eigenvalues $\lambda_{i}$ (where $i$ denotes mode number) and b), d) cumulative eigenspectrum $\sum_{j=1}^{i}\lambda_j$ normalized with the cumulative sum of the eigenvalues $\sum_{j=1}^{N_m}\lambda_j$ of each case. The colour refers to the wall-normal locations described in \S\ref{datasets}, where darker colors indicate larger distance from the wall. Note that a) and b) correspond to $Re_{\tau} = 180$, while c) and d) correspond to $Re_{\tau} = 550$. The solid black vertical lines in b) and d) refer to the number of modes selected for prediction in this study.}
    \label{fig:uc3m01}
\end{figure}

While in the work by \citet{Guemes2019sensing} the domain employed and reconstructed had a size of $h\times h$, resulting in a compact POD eigenspectrum, the availability of a larger domain in the streamwise and spanwise directions spreads the energy content over a wider set of wave numbers and POD modes (it must be recalled here that POD and Fourier modes coincide for homogeneous fields). To address this issue in the present study, the large instantaneous flow fields were subdivided into $N_{s_x}\times N_{s_z}$ smaller regions (henceforth referred to as subdomains), roughly of size $(h\times0.5h)$ in the streamwise and spanwise directions respectively for the $Re_{\tau}=180$ case, and of size $(0.4h\times0.2h)$ at $Re_{\tau}=550$. Note that the size of these subdomains is comparable to that employed by \citet{Guemes2019sensing}. 
The advantage of this approach compared to directly decomposing the full field lies in the fact that, in these subdomains, the first POD modes contain a very large fraction of the total energy content, as shown in figures~\ref{fig:uc3m01}a) and c). This is a direct consequence of including the energy of the structures larger than the domain into the first POD mode~\citep{liu2001large,wu2014study}. 
The choice of the size of the subdomains was the result of a compromise between reconstructing the majority of the energy content of the flow and compression of the information.
For $Re_{\tau}=180$ the flow fields were divided into $12\times12$ subdomains, while for $Re_{\tau}=550$ a discretization into $32\times32$ subdomains was performed. 
Note that the number of subdomains was selected to ensure that the first POD mode contains a similar level of energy in both cases.

The POD modes of the data discretized into subdomains were computed following the snapshot approach proposed by \citet{sirovich1987turbulence}.
The three fluctuating components of each instantaneous subdomain were rearranged into a snapshot matrix: 
\begin{equation}
    \mathbf{U}=\begin{bmatrix} 
    u_{x_1}^{t_1} & \dots  & u_{x_{N_p}}^{t_1} & v_{x_1}^{t_1} & \dots  & v_{x_{N_p}}^{t_1} & w_{x_1}^{t_1} & \dots  & w_{x_{N_p}}^{t_1}\\
    \vdots         & \ddots & \vdots         & \vdots         & \ddots & \vdots  & \vdots         & \ddots & \vdots \\
    u_{x_1}^{t_{N_t}} & \dots  & u_{x_{N_p}}^{t_{N_t}} & v_{x_1}^{t_{N_t}} & \dots  & v_{x_{N_p}}^{t_{N_t}}  & w_{x_1}^{t_{N_t}} & \dots  & w_{x_{N_p}}^{t_{N_t}} 
    \end{bmatrix},
\end{equation}{}
where $N_t$ refers to the total number of snapshots, {\it i.e.} equal to the number of instantaneous flow fields $N_f$ times the number of subdomains per each flow field ($N_{s_x}\times N_{s_z}\times N_f$), and $N_p$ refers to the total number of grid points in one subdomain. Then, POD spatial modes can be evaluated solving the eigenvalue problem of the spatial correlation matrix $\mathbf{C}$ as follows:
\begin{equation}
    \mathbf{C}=\mathbf{U}^T\mathbf{U}=\boldsymbol{\Phi}^T\boldsymbol{\Lambda}\boldsymbol{\Phi},
    \label{pod}
\end{equation}
where $\boldsymbol{\Phi}$ is a matrix the rows of which contain the spatial POD modes, while $\boldsymbol{\Lambda}$ is a diagonal matrix with elements $\lambda_i=\sigma_i^2$, which represent the variance content of each mode. The POD coefficients $a_i(t)$ are obtained by projecting the flow fields on the spatial POD modes computed with equation (\ref{pod}). Note that this economy-size decomposition returns a number of POD modes $N_m$ equal to $3N_p$ and that for such a discrete dataset equation (\ref{eq:decom}) is an equality.

\begin{figure}
    \centerline{\includegraphics[width=384pt]{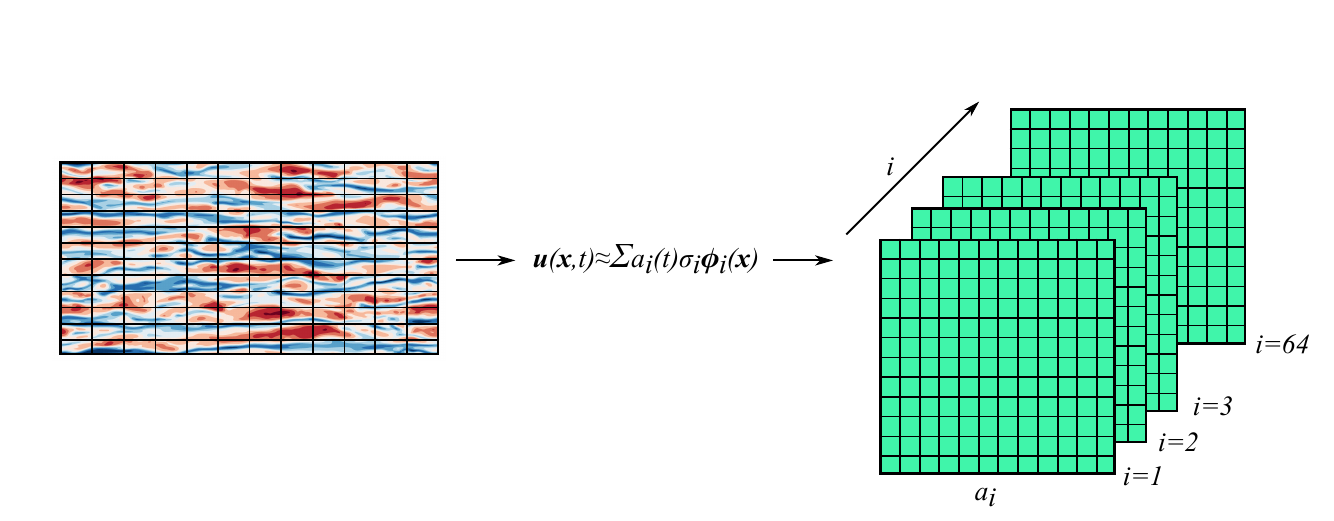}}
    \caption{Schematic representation of the encoding of turbulent flow fields into tensors containing their temporal POD coefficients.}
    \label{fig:uc3m02}
\end{figure}

The temporal POD coefficients of each instantaneous flow field were rearranged in a tensor of size $N_{s_x}\times N_{s_z}\times N_r$ to train a FCN, with $N_{s_x}\times N_{s_z}$ being $12\times12$ and $32\times32$ for $Re_{\tau} = 180$ and 550, respectively and $N_r$ the number POD modes to be predicted, with $N_{r}<N_{m}$. As shown in figures~\ref{fig:uc3m01}b) and d), the first 64 POD modes account for $90\%$ of the total energy at $Re_{\tau}=180$, while 128 modes are needed at $Re_{\tau}=550$ to retain a similar amount of energy. Therefore our predictions are based on equation (\ref{eq:decom}) truncated at $N_{r}$, with $N_{r}=64$ and $128$ for $Re_{\tau}=180$ and $550$ respectively. As illustrated in figure~\ref{fig:uc3m02}, each filter corresponds to the $N_{s_x}\times N_{s_z}$ POD coefficients of a given mode number. In general, the energy distribution reported for both $Re_{\tau}$ cases is very similar. The main significant difference is that the energy distribution becomes more compact at $y^+=15$ for the low-$Re_{\tau}$ case (see figure \ref{fig:uc3m01}).

In order to reconstruct the instantaneous fluctuation fields, the time coefficients were predicted using a FCN, with the wall-shear-stress components and the wall pressure as inputs. The $N_r$ time coefficients belonging to each subdomain are used to reconstruct their respective fluctuation fields as in equation~(\ref{eq:decom}), where the orthonormal basis functions are retrieved from the training data. The entire fields are assembled by tiling the fields within these subdomains. Note that there is no guarantee of smoothness across the edges of the subdomains because of the finite number of modes that are used to reconstruct the flow and because of the prediction error in the temporal coefficients.
The underlying assumption is \textit{ergodicity}, {\it i.e.} both the training and test datasets share the same statistical features and, consequently, the same spatial modes. This requires a sufficiently large training dataset to ensure convergence of the spatial modes, which is generally ascribed to the convergence of second-order statistics. Note that the predictions are performed at the same instant as that of the input fields. The implemented neural network does not require the knowledge of the input at previous timesteps, thus avoiding the limitations of availability of time sequences, as discussed above. The network is trained to minimize the loss function: 
\begin{equation}
\mathcal{L}_{{\rm FCN-POD}}(a_\mathrm{POD};a_\mathrm{DNS})=\frac{1}{N_{s_x} N_{s_z} N_r} \sum_{i=1}^{N_{s_x}}  \sum_{j=1}^{N_{s_z}} \sum_{k=1}^{N_r} \left | a_\mathrm{POD}(i,j,k) -a_\mathrm{DNS}(i,j,k)\right |^{2},
\end{equation}
which is the MSE between the predicted and the actual POD temporal coefficients of the DNS data.

The neural-network architecture considered here blends the FCN shown in figure~\ref{fig:net} and the network used by \citet{Guemes2019sensing} (see figure~1 in that work). As in the FCN approach, each convolution operation (except for the last one) is followed by batch normalization~\citep{ioffe} and a ReLU~\citep{nair2010rectified} activation function. After each activation function a max pooling layer is added. Differently from what was done in the FCN approach, here the velocity components were not scaled before the decomposition, in order to keep the physical encoding based on the turbulent kinetic energy (TKE) of the flow. Note that by modifying the relative contribution of the velocity components to the energy norm, the modes would have been sorted based on a norm different from the TKE.
The main difference with respect to the network used by \citet{Guemes2019sensing} is the fact that here a single network is used to predict the full set of POD coefficients, instead of using different networks to predict each mode. Additionally, the work by \citet{Guemes2019sensing} focused on a smaller region of the flow field, and therefore no subdomains were required for the region of interest. Lastly, the final fully-connected layer in \citet{Guemes2019sensing} was not considered here, in order to have an architecture more directly comparable with the FCN. As in the FCN case, the FCN-POD network was trained using Adam~\citep{kingmaba} optimization algorithm with a scheduled exponential learning-rate decay. In this case, the $\hat{\epsilon}$ parameter from \citet{kingmaba} was set to 0.1 following TensorFlow recommendations~\citep{abadi2016tensorflow}. For the $Re_{\tau}=180$ case the number of trainable parameters is 4,733,248, while for the $Re_{\tau}=550$ case it is 5,028,224. 

\subsection{Extended POD}\label{ss:epod}
In addition to the two FCN-based approaches, which involve nonlinear relations between input and output, we also consider a method involving a linear relationship, {\it i.e.} the EPOD. Doing so, it will be possible to assess the prediction improvement with nonlinear methods in the context of wall-bounded turbulence. If the wall quantities are rearranged into a snapshot matrix $\mathbf{W}$, with each snapshot forming a row, the method of snapshots proposed by \citet{sirovich1987turbulence} can be used to decompose this matrix into POD modes as:

\begin{equation}
    \mathbf{W}=\boldsymbol{\Psi}_w\boldsymbol{\Sigma}_w\boldsymbol{\Phi}_w,
    \label{wpod0}
\end{equation}
with $\boldsymbol{\Psi}_w$ and $\boldsymbol{\Phi}_w$ being the temporal and spatial mode matrices respectively, and $\boldsymbol{\Sigma}_w$ being a diagonal matrix containing the singular values. The extended POD modes~\citep{boree2003extended}, corresponding to the projection of the wall quantities on the flow-field temporal basis, are defined as:
\begin{equation}
    \mathbf{L}=\boldsymbol{\Psi}_w^T\mathbf{U}.
    \label{wpod1}
\end{equation}
If the dataset is sufficiently large to reach statistical convergence, the matrix $\mathbf{L}$ describes the relationship between the temporal POD coefficients of a certain distribution of wall features, and those of the corresponding flow field. Once the temporal correlation matrix is known, an out-of-sample flow field $\mathbf{u}$ can be reconstructed using $\mathbf{L}$ and the instantaneous realization of wall features as follows: 
\begin{equation}
    \mathbf{u}=\boldsymbol{\psi}_{w}\mathbf{L},
    \label{epod02}
\end{equation}
where $\boldsymbol{\psi}_{w}$ is the vector containing the temporal coefficients of the wall fields used for prediction. Note that $\boldsymbol{\psi}_{w}$ is retrieved by projecting the out-of-sample wall field $\boldsymbol{w}$ on the POD basis: $\boldsymbol{\psi}_{w}=\boldsymbol{w}\boldsymbol{\Phi}_{w}^T \boldsymbol{\Sigma}_{w}^{-1}$, where $\boldsymbol{\Phi}_{w}^T \boldsymbol{\Sigma}_{w}^{-1}$ is readily available from the training dataset. 

An important remark is that the matrix $\boldsymbol{\Sigma}_{w}$ can be ill-conditioned. In fact due to the correlation between subsequent time-resolved snapshots, the rank of $\boldsymbol{\Sigma}_{w}$ is smaller than $N_f$, which is the number of snapshots (here smaller than the number of points). To address this issue a reduced-order representation of the matrix $\boldsymbol{\Sigma}_w$ is employed after truncating the null elements in the diagonal of $\boldsymbol{\Sigma}_w$.
Even if $\boldsymbol{\Sigma}_w$ would have rank equal to $N_f$, it might be adequate to truncate the matrix $\boldsymbol{L}$  \citep{discetti2018estimation}. Decomposing the flow quantities as: $\mathbf{U}=\boldsymbol{\Psi}_{u}\boldsymbol{\Sigma}_{u}\boldsymbol{\Phi}_{u}$, similarly to what is done for the wall quantities in equation~(\ref{wpod0}), it can be observed that the product of the two matrices  $\boldsymbol{\Psi}_{u}\boldsymbol{\Psi}_{w}^{T}$ in equation (\ref{wpod1}) returns a unitary-norm matrix with rank equal to those of $\boldsymbol{\Psi}_{u}$ and $\boldsymbol{\Psi}_{w}^{T}$, which are bases in the $\mathbb{R}^{N_f}$ vector space. As a consequence, a certain $j^{th}$ wall mode, uncorrelated with any field mode, would not result in a corresponding null row or column. To ensure removing the uncorrelated content from the matrix $\boldsymbol{\Psi}_{u}\boldsymbol{\Psi}_{w}^{T}$, \citet{discetti2018estimation} proposed to set to zero all the entries of the matrix with absolute values smaller than a threshold proportional to the matrix standard deviation. In the present work we have found an error drop of approximately 10 percentage points with respect to the standard EPOD procedure. However, since the EPOD is used as a benchmark for the performance of linear methods with respect to the FCN-based approaches proposed herein, the filtered EPOD by \citet{discetti2018estimation} is not included in this comparison for brevity.

\section{Results}\label{ss:results}
The predictions of the trained models are compared with the data obtained from the DNS at $Re_{\tau}=180$ and $550$. The performance assessment is carried out first from a qualitative point of view and subsequently from a quantitative perspective, based on predictions of instantaneous fields, turbulence statistics and spectra.
\subsection{Instantaneous predictions}\label{ss:inst}
\begin{figure}
\begin{center}
\includegraphics[width=\textwidth]{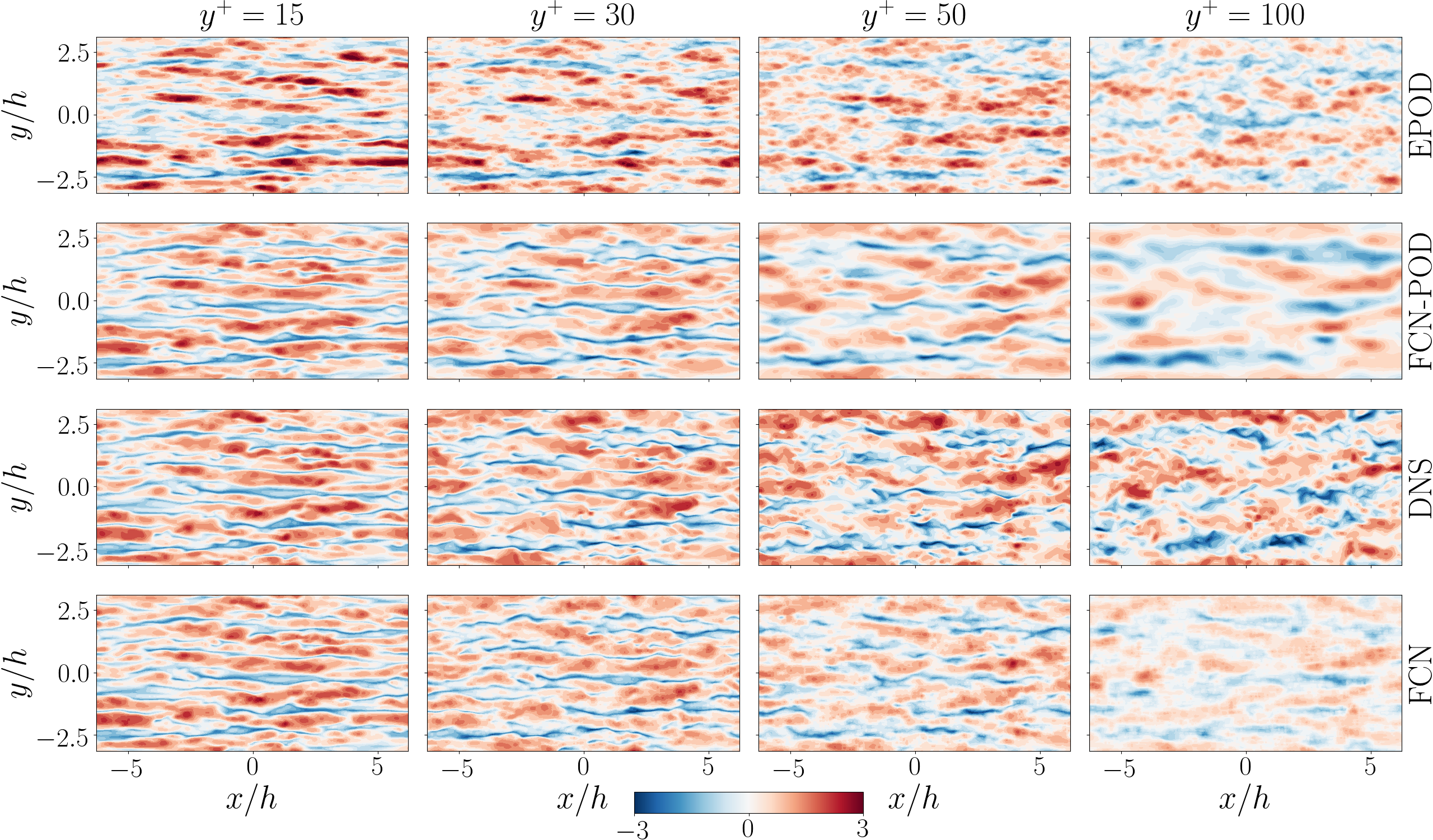}
\end{center}
\caption{\label{fig:field_comp180} Comparison of the streamwise fluctuation fields at $Re_{\tau} = 180$, scaled with the corresponding $u_\mathrm{RMS}$, from EPOD (1$^{\text{st}}$ row), FCN-POD (2$^{\text{nd}}$ row), reference DNS (3$^{\text{rd}}$ row) and FCN (4$^{\text{th}}$ row). Results at $y^+=15$ (1$^{\text{st}}$ column), $y^+=30$ (2$^{\text{nd}}$ column), $y^+=50$ (3$^{\text{rd}}$ column) and $y^+=100$ (4$^{\text{th}}$ column).}
\end{figure}

The predicted fluctuation fields are first qualitatively inspected. Note that the fluctuation flow fields are the direct output of the FCN models, while in the FCN-POD models the temporal coefficients need to be processed to reconstruct the fluctuations, as outlined above. In this work the sampling period in the simulation is fixed, however we showed in our previous work~\citep{guastoni} that using less correlated samples during training (\textit{i.e.} higher sampling period) can effectively improve the quality of the instantaneous predictions of the FCN method, provided that the neural-network capacity is sufficient to generalize over the training dataset. 

In figure~\ref{fig:field_comp180}, the predictions of an instantaneous field of streamwise velocity fluctuations based on the various methods (namely FCN, FCN-POD and EPOD) are compared with the reference DNS. The predictions of the wall-normal and spanwise fluctuations at the same instant are shown in Appendix~\ref{appA}. At $y^+=15$ all the methods provide accurate results, although the EPOD overestimates the fluctuations from the high-speed streaks. At $y^+=30$ the neural-network-based models maintain a good level of accuracy while the EPOD, despite the improved predicted range of fluctuations, it does not seem to be as accurate as the other two models. The CNN-based methods start to exhibit some deviations with respect to the reference at $y^+=50$, where the FCN-POD field is smoother and the FCN is slightly noisier than the DNS. Farther from the wall, the footprint of the large scales at the wall~\cite[through linear superposition, see][]{dogan_et_al} is less pronounced, and therefore the ability of EPOD (which is a linear method) to predict the flow in this region is significantly reduced. In fact, the fields predicted through EPOD at $y^{+}>15$ are qualitatively very similar to the DNS, although the fluctuations become increasingly attenuated at larger $y^{+}$. Furthermore, the FCN-POD method tends to merge neighbouring regions with high- or low-velocity fluctuations, predicting more elongated streak-like patterns than in the reference field. This is more evident at $y^+=100$, leading to an overestimation of the amplitude of the regions in the flow where the velocity fluctuations are higher. At this location, the FCN is not able to provide a reliable prediction of the flow field, capturing only the regions in which the magnitude of the fluctuations is higher. The corresponding structures probably have a distinct footprint at the wall, which allows the FCN to identify them. Note that at $Re_{\tau}=180$ there is no real scale separation. 

As discussed in $\S$\ref{ss:FCN-POD}, the FCN-POD approach does not guarantee flow smoothnes across the edges of the subdomains. Close inspection of the predictions from the FCN-POD method reveals the edges of the subdomains at all $y^+$, and the tiling is more evident in the streamwise direction because of the discontinuities located at the same spanwise location, orthogonally to the main flow structures. Despite these limitations we can observe that the velocity fields are generally smooth, without steep discontinuities at the edges of the subdomains: the variations of the velocity magnitude at the edges are of the same order as the fluctuations at the corresponding wall-normal distance. 

\begin{figure}
\begin{center}
\includegraphics[width=.328\textwidth]{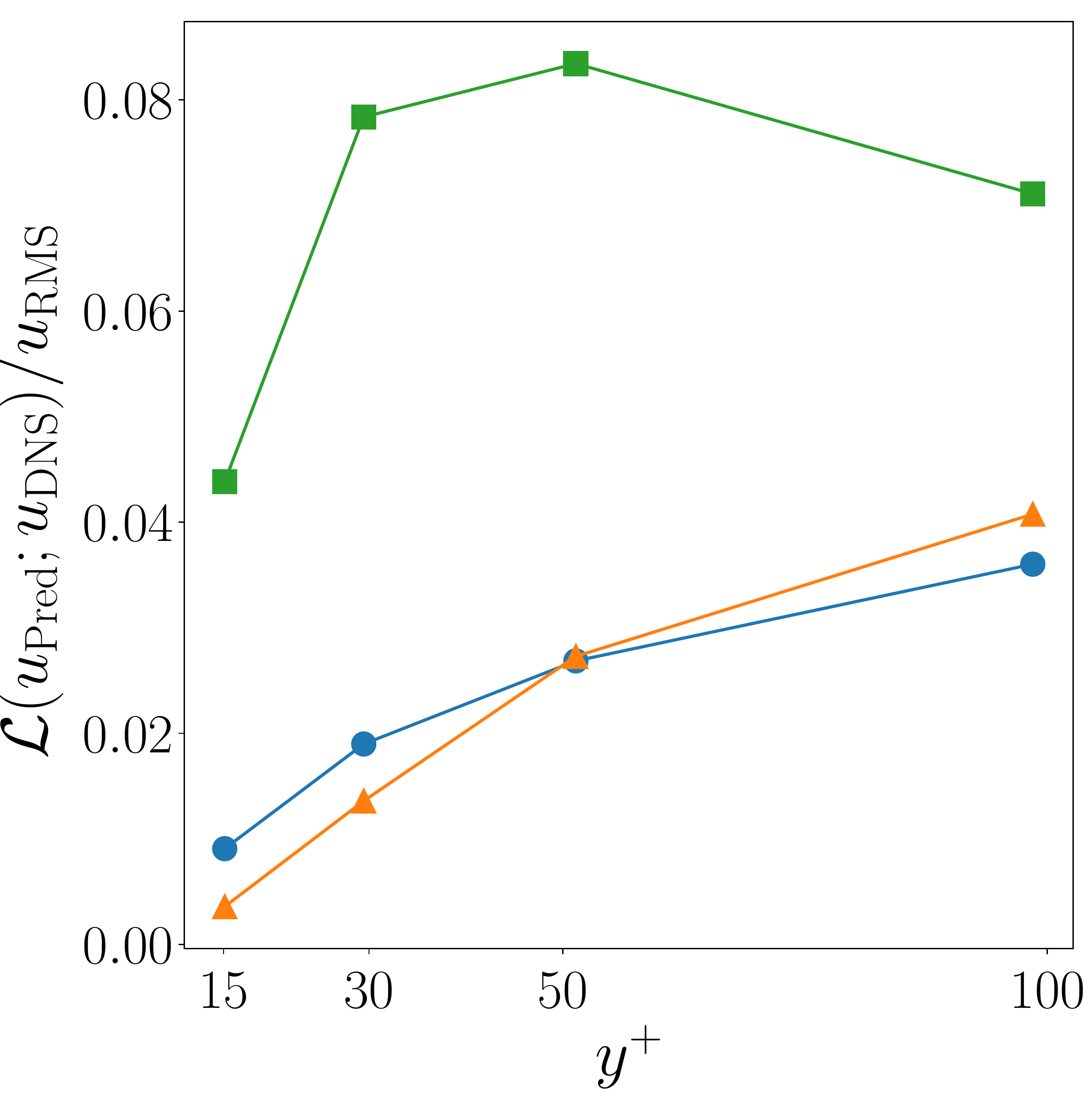}
\includegraphics[width=.328\textwidth]{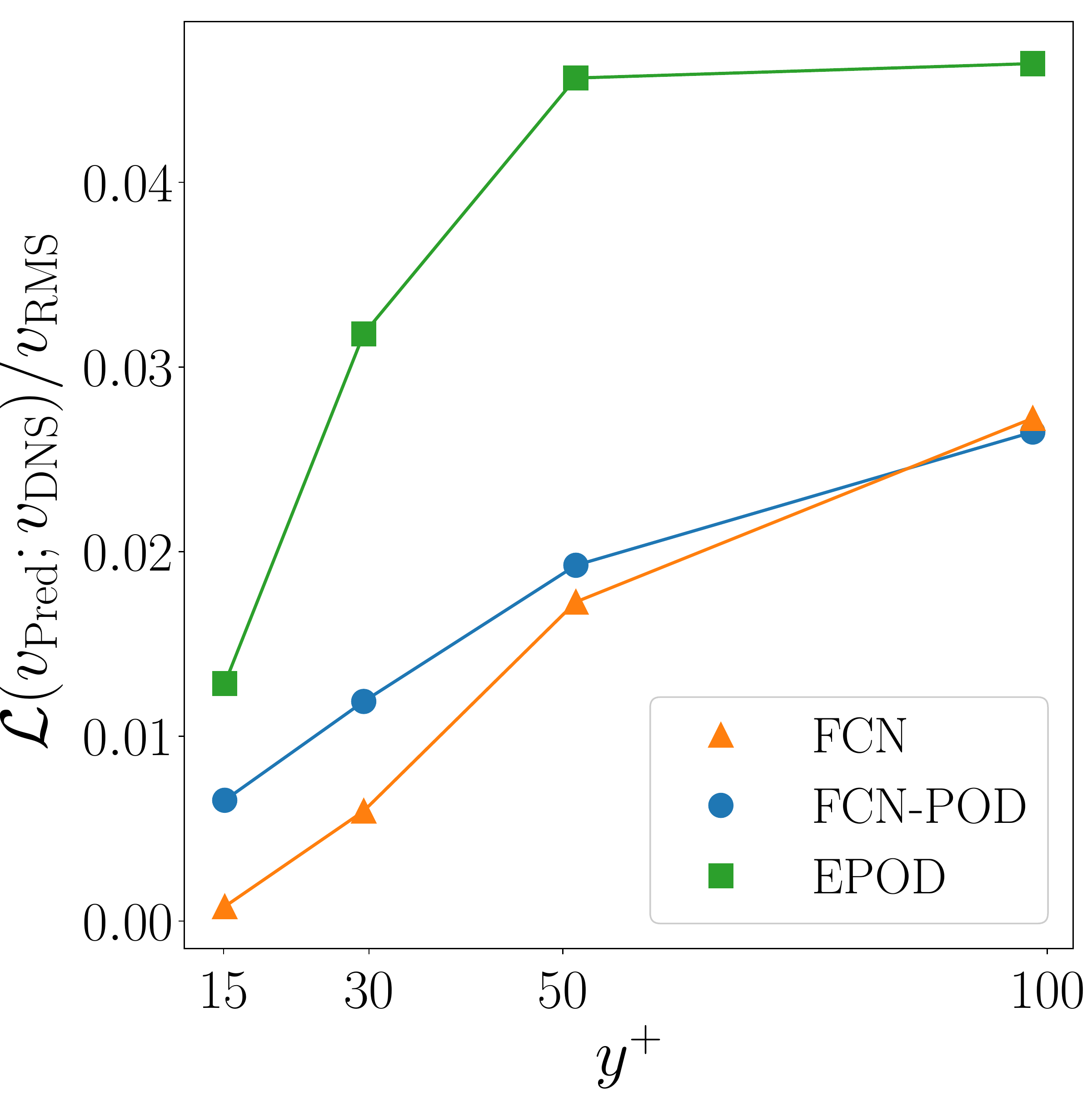}
\includegraphics[width=.328\textwidth]{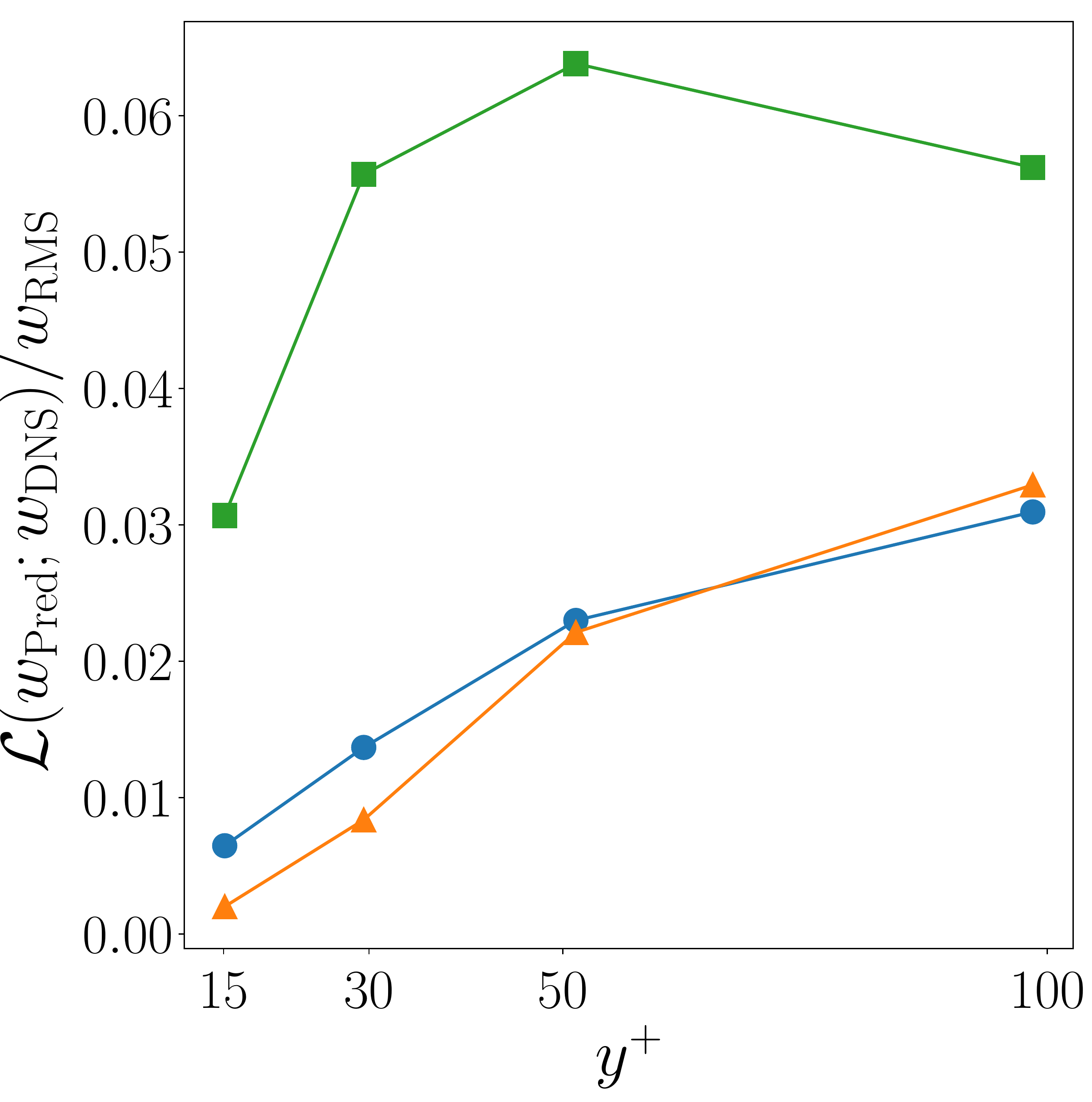}
\end{center}
\caption{\label{fig:loss180} Mean-squared-error in the instantaneous fields (scaled with the corresponding RMS components) predicted by the three models at $Re_{\tau} = 180$, for the streamwise (left), wall-normal (middle) and spanwise (right) velocity fluctuations.}
\end{figure}

The qualitative observations discussed above are complemented with a quantitative assessment of the instantaneous prediction performance, by analyzing the MSE $\mathcal{L}$ between the instantaneous predictions (denoted by `Pred') and the reference (defined for each of the fluctuations independently), as shown in figure~\ref{fig:loss180}. Both neural-network models (FCN and FCN-POD) are trained using a stochastic algorithm and, in order to show the robustness of the optimal configuration, the statistics at each $y^+$ are averaged over 3 different models, with different initial random weight initializations. Since the EPOD algorithm is completely deterministic, one single prediction is needed. The neural-network-based models consistently provide a lower error than the EPOD, with the FCN yielding a slightly better instantaneous performance closer to the wall than the FCN-POD approach. The gap between the two is reduced when moving away from the wall, where the prediction error of the streamwise fluctuations is approximately the same for both models at $y^+=50$, and it is slightly higher for the FCN at $y^+=100$. 

The FCN architecture reported by \cite{guastoni} would only predict the streamwise velocity component of the velocity field at the target $y^+$. The addition of the two other components implies that the FCN has multiple outputs that need to be optimized at the same time. We note that adding the two additional fluctuating components as outputs leads to slightly less accurate predictions with respect to those reported by \cite{guastoni} for one single output. This is not surprising, since the capacity of the network remained unchanged, however we tested a variation of the model architecture based on this observation, in order to have more layers dedicated to the prediction of each individual component. This network variation has a common part, identical to the original FCN up to the 4$^{\text{th}}$ convolutional layer, in which the weights are optimized using the information from the error gradients computed for all the outputs. The last two convolution operations are replicated for each velocity component and the weights of these layers are updated only with the error associated to the respective output. Such a network, despite its higher capacity, provided worse predictions. A strong causal relation between the different components of the velocity~\citep{lozano2020causality} can be a possible explanation for this result, which shows that updating all the weights with information from the three components at the same time can be beneficial for the quality of the predictions. Note that it is not trivial to design an architecture able to provide the best trade-off between single-component predictions and usage of the information from all the components, and obtaining such an architecture would require further investigation. For the FCN-POD model the multiple-component predictions were obtained as discussed by \citet{Guemes2019sensing}. The temporal POD coefficients can be projected on spatial POD modes involving the three velocity components, thus requiring only one output to predict the three fluctuations. The network architecture is different than the one used by \citet{Guemes2019sensing}, since it predicts directly all the needed time coefficients for each snapshot. While the final fully-connected layer included in the network architecture by \citet{Guemes2019sensing} improves the robustness of the prediction, the FCN-POD implementation used here has a much smaller number of weights, thus significantly reducing the computational cost, and retains a larger number of POD modes (and thus more energy).

\begin{figure}
\begin{center}
\includegraphics[width=\textwidth]{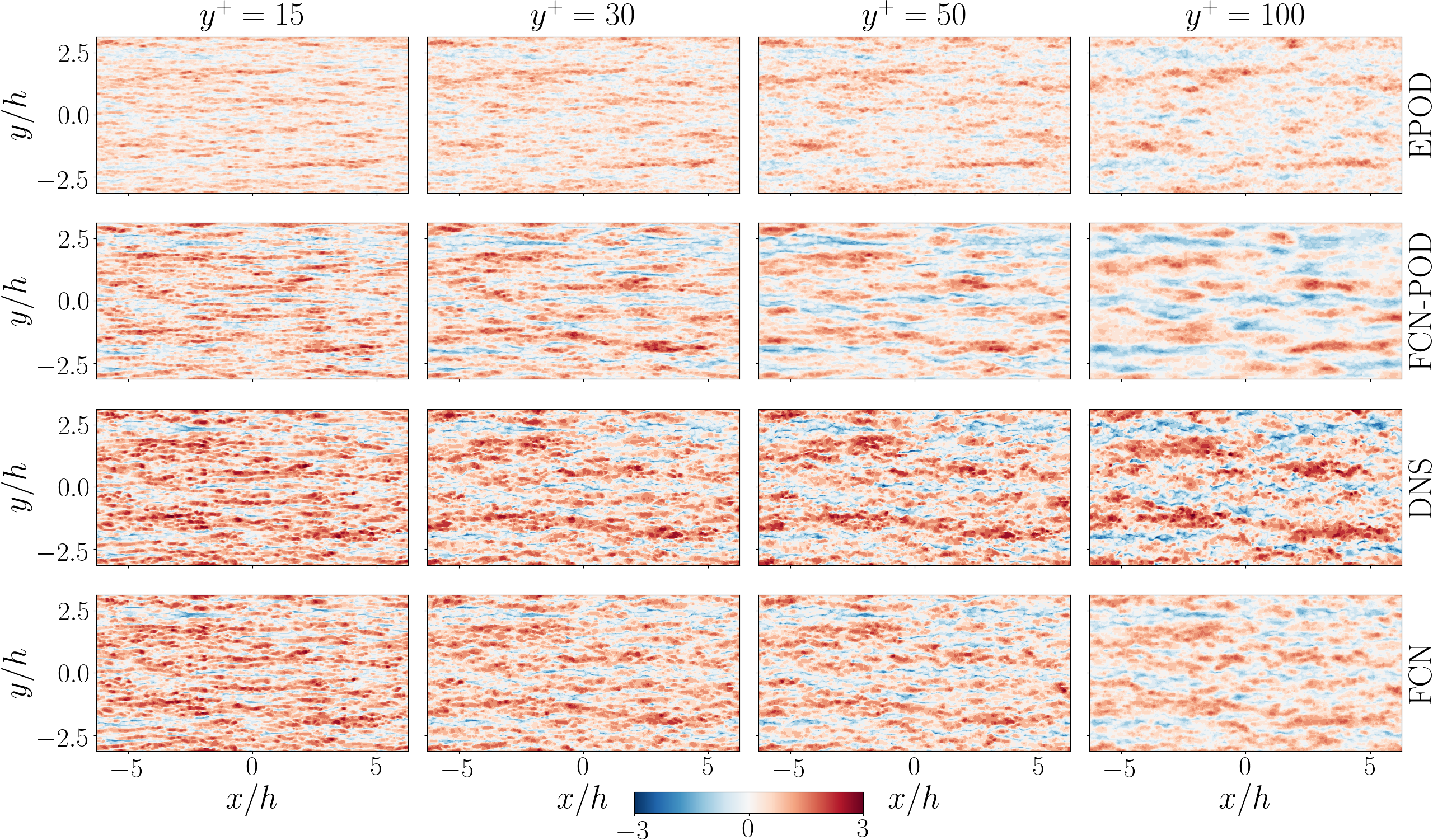}
\end{center}
\caption{\label{fig:field_comp550} Comparison of the streamwise fluctuation fields at $Re_{\tau} = 550$, scaled with the corresponding $u_\mathrm{RMS}$, from EPOD (1$^{\text{st}}$ row), FCN-POD (2$^{\text{nd}}$ row), reference DNS (3$^{\text{rd}}$ row) and FCN (4$^{\text{th}}$ row). Results at $y^+=15$ (1$^{\text{st}}$ column), $y^+=30$ (2$^{\text{nd}}$ column), $y^+=50$ (3$^{\text{rd}}$ column) and $y^+=100$ (4$^{\text{th}}$ column).}
\end{figure}

\begin{figure}
\begin{center}
\includegraphics[width=.328\textwidth]{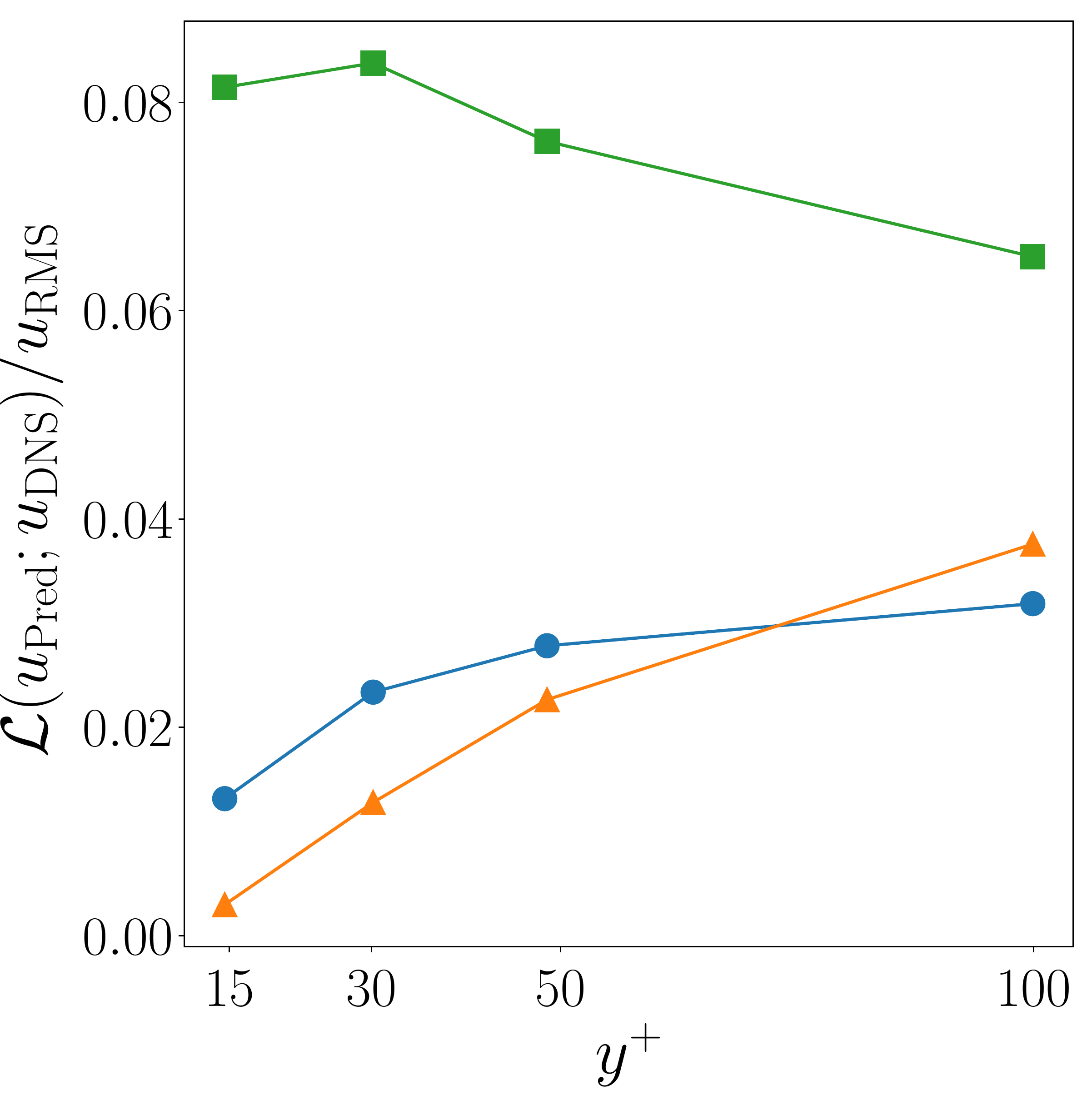}
\includegraphics[width=.328\textwidth]{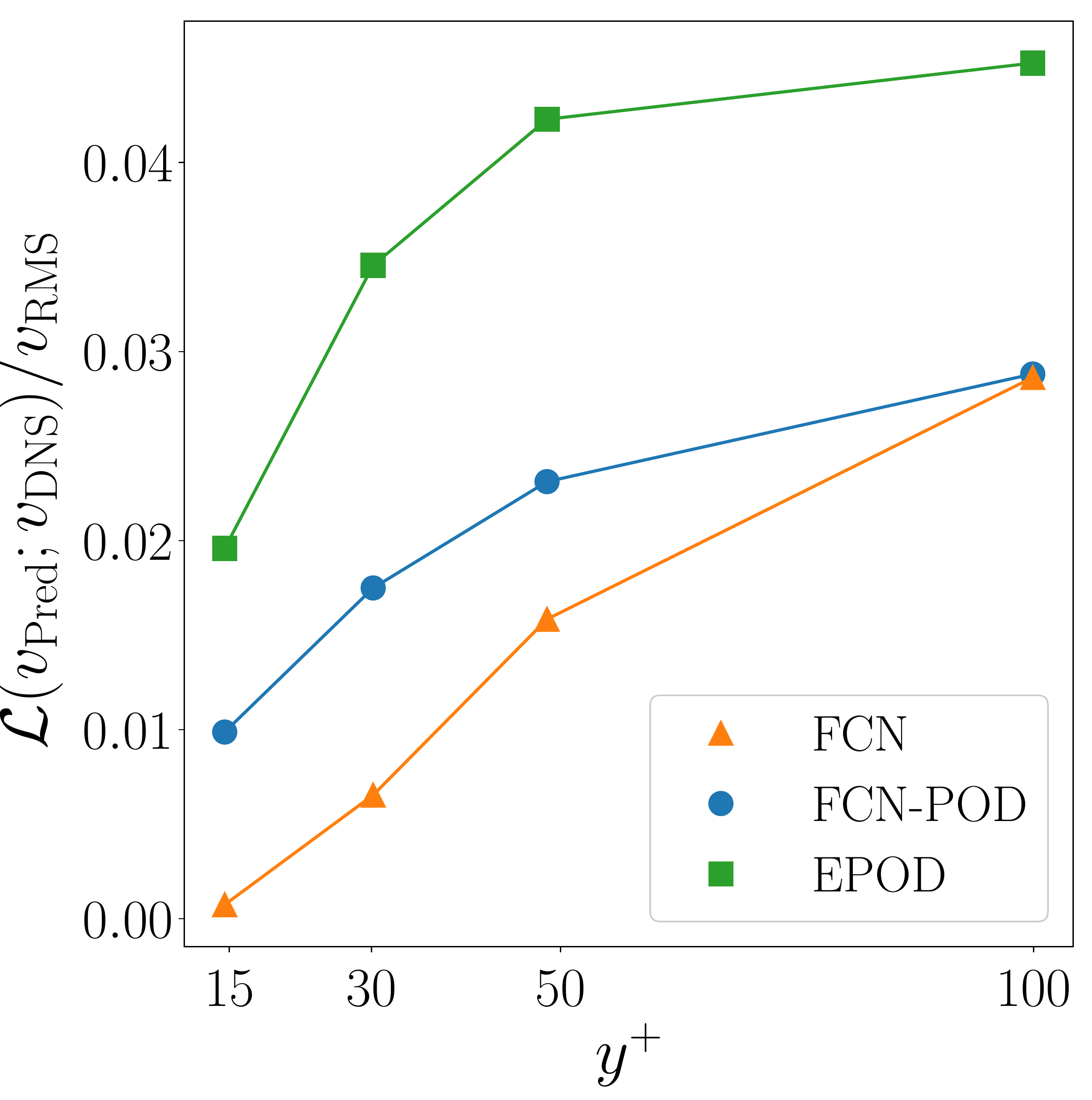}
\includegraphics[width=.328\textwidth]{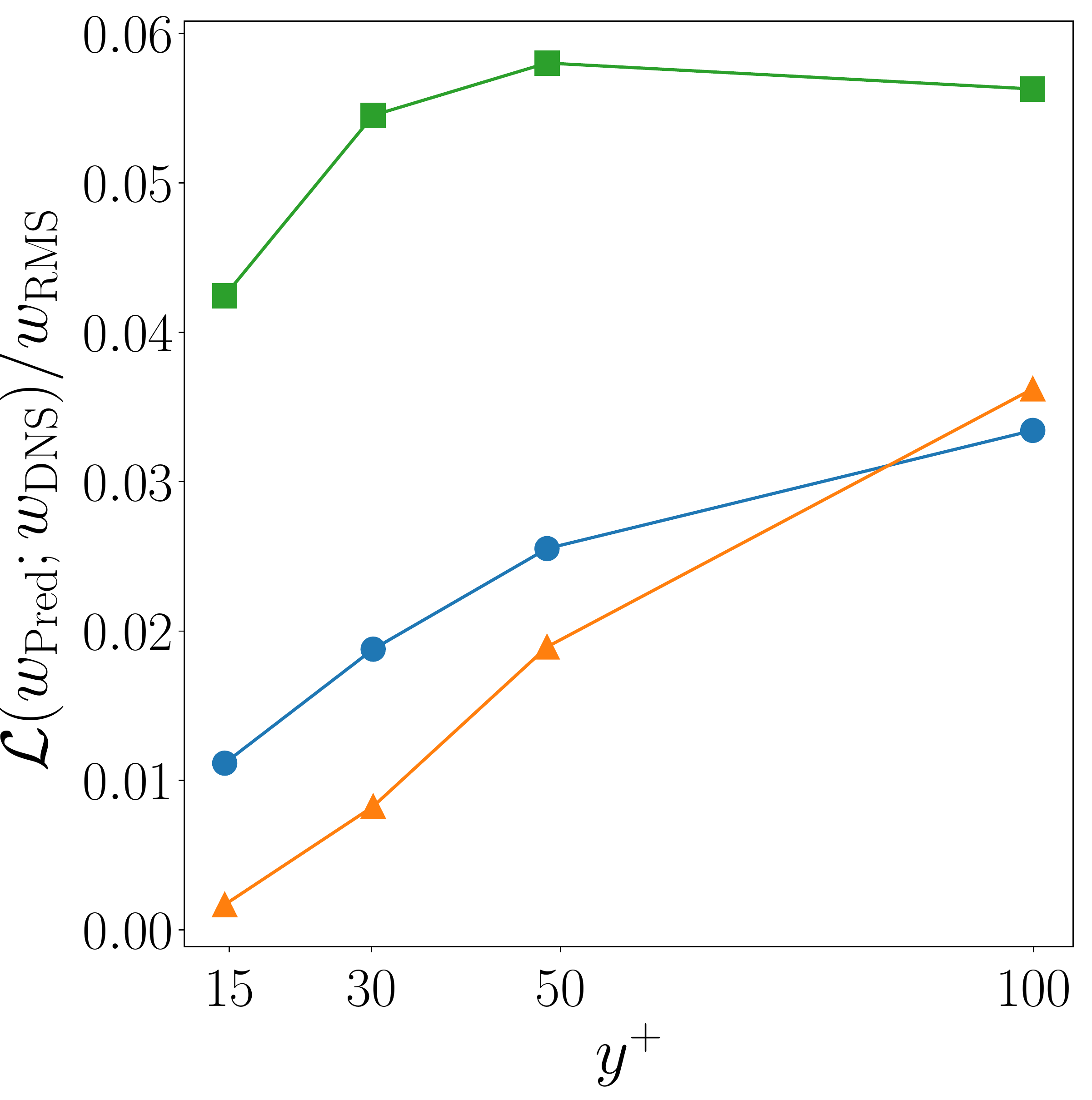}
\end{center}
\caption{\label{fig:loss550} Mean-squared-error in the instantaneous fields (scaled with the corresponding RMS components) predicted by the three models at $Re_{\tau}=550$, for the streamwise (left), wall-normal (middle) and spanwise (right) velocity fluctuations.}
\end{figure}

Predictions of the streamwise fluctuation fields from the various methods at $Re_{\tau} = 550$ are shown in figure~\ref{fig:field_comp550}, while the results for the wall-normal and spanwise components are presented in Appendix~\ref{appA}. Despite the higher friction Reynolds number, the FCN maintains a performance similar to the one achieved for $Re_{\tau}=180$, at all wall-normal locations. Note that the FCN has the same architecture as the lower-Reynolds-number case, {\it i.e.} it has the same number of trainable parameters, while in the case of the FCN-POD approach, the network was modified to reconstruct approximately the same amount of energy as at $Re_{\tau}=180$. Despite the higher number of employed subdomains, the tiling is more apparent at $Re_{\tau}=550$. The prediction performance of the FCN-POD model degrades less quickly than the FCN when moving away from the wall, however the latter still performs better at $y^+=50$, as shown in figure~\ref{fig:loss550}. On the other hand, the EPOD also exhibits similar error levels as those reported for $Re_{\tau}=180$, except at $y^+=15$, where the reconstruction of the streamwise-fluctuation field is significantly worse. 

\subsubsection{Inclination of coherent structures} \label{sss:shift}
The coherent structures in wall-bounded turbulence are inclined~\citep{marusic_heuer}, with a slope that can be computed by finding the maximum spatial correlation $\mathcal{R}_{ij}(\delta x)$ between the inputs at the wall (index $i$) and the outputs (index $j$), with $\delta x$ representing the distance in the streamwise direction at which the correlation is computed. By including a streamwise \textit{shift} in the output fields, it is possible to obtain the maximum correlation at $\delta x=0$, ensuring that the footprint of the coherent structure is included in the receptive field of the output. The use of such a shift was also discussed by \citet{sasaki_vinuesa_cavalieri_schlatter_henningson_2019} in a similar context. By considering the maximum correlation between the wall-shear stress in the streamwise direction and the streamwise velocity at a certain $y^+$, we obtain an angle of $\approx 15^{\circ}$, in very good agreement with previous observations~\citep{marusic_heuer,sasaki_vinuesa_cavalieri_schlatter_henningson_2019}. This shift was implemented in two alternative ways: first, by modifying the target output field, {\it i.e.} considering a field that has been sampled later in the simulation, although the accuracy of the introduced shift is limited by the value of the sampling period. The second approach makes use of the periodicity of the output fields, which are translated in the streamwise direction until the maximum correlation is obtained at $\delta x=0$. This approach allows to more accurately introduce the shift, however in this case the underlying hypothesis is that the shift is sufficiently small so that temporal dynamics modify the flow in a negligible manner. None of the two shift implementations provided the expected improvement, and we observed a significant degradation of the prediction performance. These results could possibly be explained by the fact that coherent structures of different size have different inclinations, and imposing a single value is detrimental for the overall network performance, despite having chosen the angle that provides the maximum spatial correlation. Furthermore, the quality of the predictions is measured using the MSE between the prediction and the reference: this error indicator considers all wavelengths at the same time, without considering how the different wavelengths are affected by the shift. Further investigation of this aspect will be conducted in future work.

\subsection{Predictions of turbulence statistics}\label{ss:stat}
\begin{figure}
\begin{center}
\includegraphics[width=.328\textwidth]{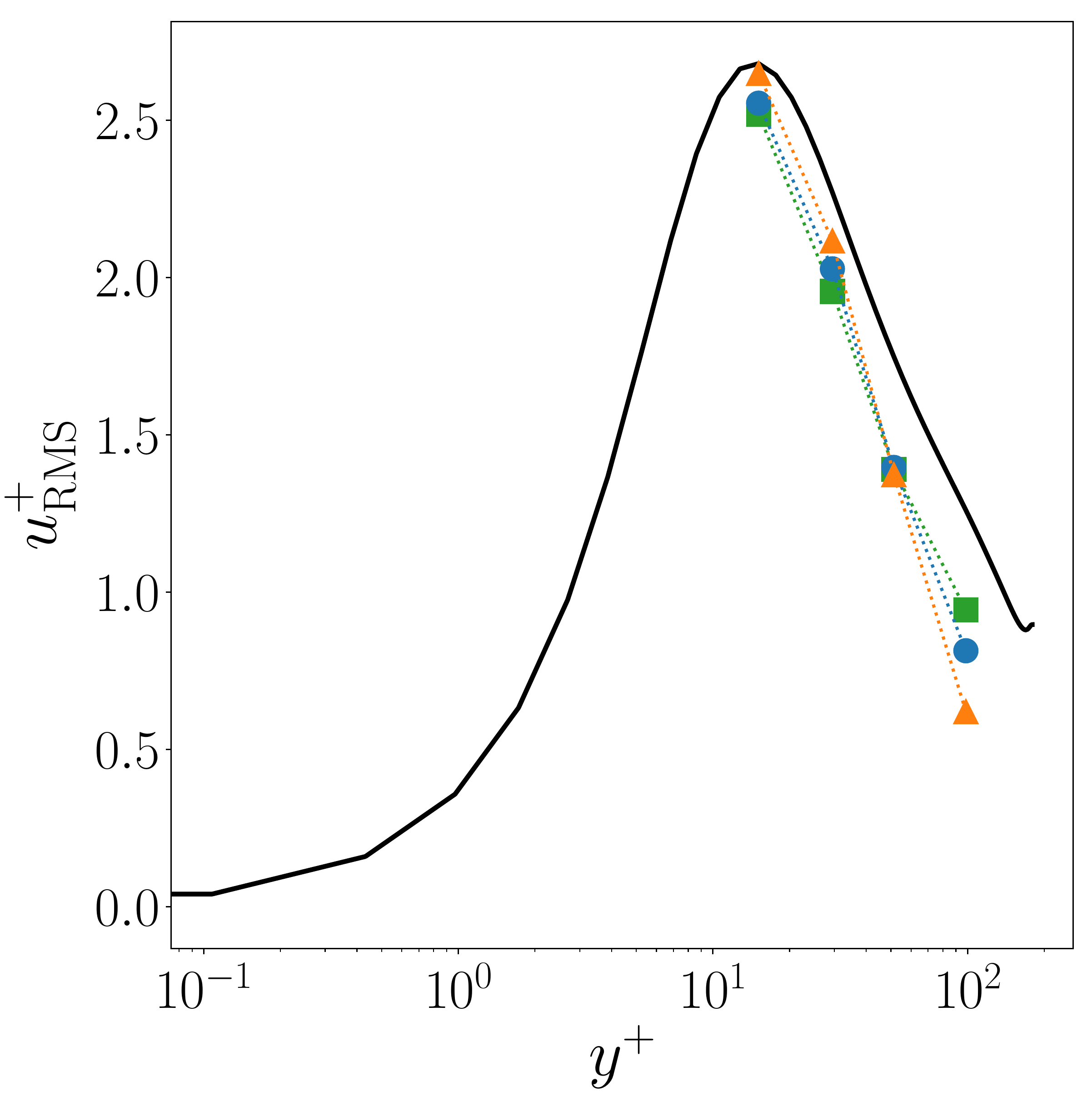}
\includegraphics[width=.328\textwidth]{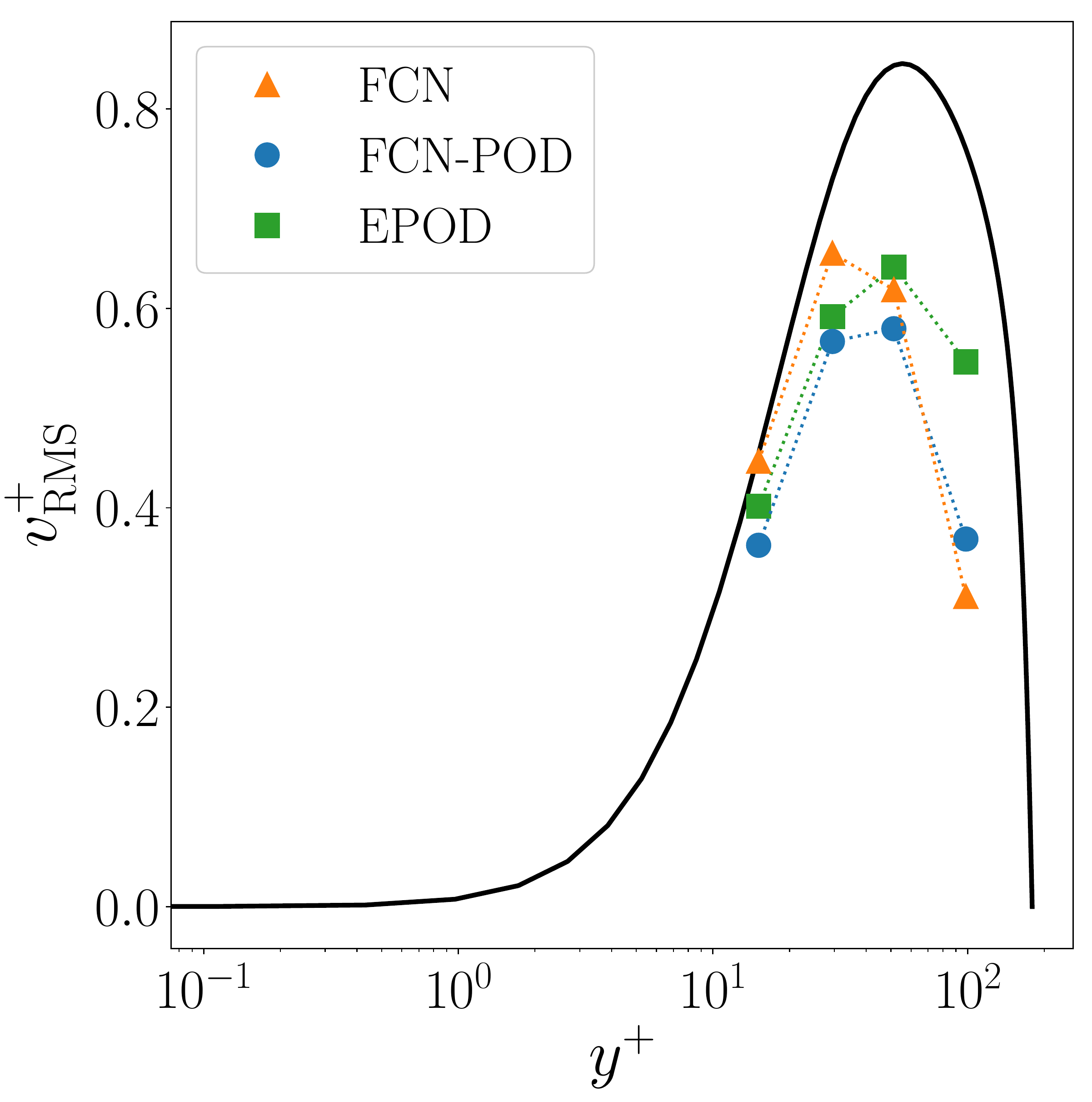}
\includegraphics[width=.328\textwidth]{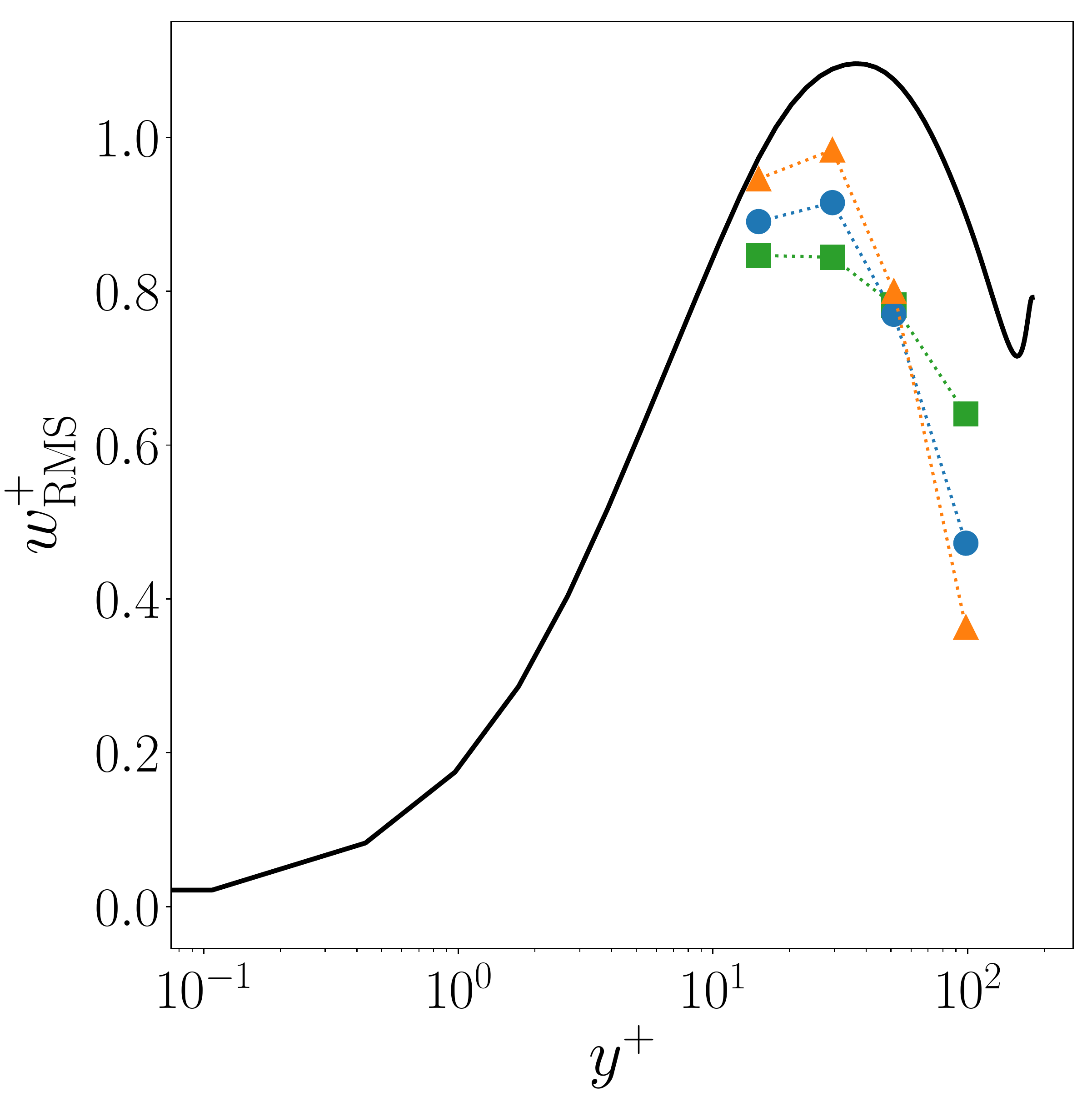}
\end{center}
\caption{\label{fig:stats180} Comparison between the DNS ( \full) and the predictions of streamwise (left),
 wall-normal (middle) and spanwise (right) velocity 
fluctuations at $Re_{\tau} = 180$.}
\end{figure}

By averaging over the fields obtained from the neural-network models and EPOD, it is possible to evaluate the turbulence statistics of the predicted flow. First we consider the dataset at $Re_{\tau}=180$: the predicted RMS fluctuations of the three components are shown in figure~\ref{fig:stats180}, together with the reference DNS profiles. The error in these statistical quantities is defined as:
\begin{equation}
    E_\mathrm{RMS}^+(u) = \frac{\left| u_\mathrm{RMS,Pred}^+ - u_\mathrm{RMS,DNS}^+ \right|}{u_\mathrm{RMS,DNS}^+},
\end{equation}
for the streamwise component, and similarly for the other two components. As above, the subscripts `DNS' and `Pred' refer to the reference and predicted profiles, respectively. An important premise is that neither of the neural-network-based models is explicitly optimized to reproduce the statistics of the original simulation. This prevents the neural networks from learning only the average behaviour of the flow, however the predictions may be less statistically accurate, with the aim of maximizing the instantaneous performance. Note that here we favor instantaneous performance because our motivation is to use non-intrusive sensing for closed-loop flow control. 

The prediction errors in the various RMS profiles are summarized in table~\ref{tab:RMS_err_180}, and they are averaged over the different training runs for the FCN and FCN-POD models. Note that the average is performed over the fluctuation-intensity values and not on the predictions, because that would alter the statistical properties of the predicted flow fields. The comparison of the errors from the different models shows that the statistical performance mimics the one observed for the instantaneous predictions at $y^+=15$ and $y^+=30$, with the FCN performing better than the FCN-POD and EPOD models. Furthermore, the FCN model provides a similar performance for the fluctuations of all three velocity components, while POD-based methods are more accurate in the predictions of $u_\mathrm{RMS}^+$. This is related to the choice of not scaling the different velocity components in the FCN-POD and EPOD approaches, and the fact that near the wall the most energetic dynamics of the flow are in the streamwise direction. 
Taking into account the standard deviation in the results of the neural-network-based methods, the three models provide similar error levels at $y^+=50$. At $y^+=100$ the scenario is opposite to what we observed close to the wall: the FCN exhibits the highest errors, while the EPOD provides the best results. The error in the prediction of $u_\mathrm{RMS}^+$ from the FCN-POD model is between those of the two other models, while the wall-normal and spanwise intensities are closer to the errors from the FCN, due the reasons outlined above. 

\begin{table}
\begin{center}
    \begin{tabular}{l l*{4}{c}}
        $E_\mathrm{RMS}^+(\cdot)\ [\%]$ &         & $y^+=15$                       & $y^+=30$                     & $y^+=50$           & $y^+=100$          \\[0.3cm]
        
                                        & EPOD    & $\phantom{0}6.03~~~~~~~~~~~$   & $13.87~~~~~~~~~~~$           & $20.50~~~~~~~~~~~~$ & $25.15~~~~~~~~~~~~$ \\
        $u$                             & FCN     & $\phantom{0}1.16~(\pm 0.74)$   & $\phantom{0}6.79~(\pm 1.31)$ & $21.47~(\pm 1.97)~$ & $50.82~(\pm 2.19)~$ \\
                                        & FCN-POD & $\phantom{0}4.70~(\pm 0.02)$   & $10.70~(\pm 0.02)$           & $20.15~(\pm 0.03)~$ & $35.46~(\pm 0.02)~$ \\[0.3cm]
                                     
                                        & EPOD    & $11.68~~~~~~~~~~~$             & $18.89~~~~~~~~~~~$           & $23.97~~~~~~~~~~~~$ & $28.10~~~~~~~~~~~~$ \\
        $v$                             & FCN     & $\phantom{0}1.74~(\pm 1.00)$   & $10.18~(\pm 1.67)$           & $26.66~(\pm 0.76)~$ & $59.05~(\pm 1.61)~$ \\
                                        & FCN-POD & $20.29~(\pm 0.02)$             & $22.32~(\pm 0.02)$           & $31.32~(\pm 0.01)~$ & $51.48~(\pm 0.04)~$ \\[0.3cm]
                                     
                                        & EPOD    & $13.01~~~~~~~~~~~$             & $22.48~~~~~~~~~~~$           & $27.27~~~~~~~~~~~~$  & $28.72~~~~~~~~~~~~$ \\
        $w$                             & FCN     & $\phantom{0}2.79~(\pm 0.36)$   & $\phantom{0}9.65~(\pm 1.07)$ & $25.60~(\pm 1.214)$  & $59.59~(\pm 1.310)$ \\
                                        & FCN-POD & $\phantom{0}8.50~(\pm 0.04)$   & $15.95~(\pm 0.06)$           & $28.38~(\pm 0.004)$  & $47.42~(\pm 0.001)$ \\                             
    \end{tabular}
    \caption{Percentage error in the prediction of the various RMS fluctations at the different wall-normal locations. Results at $Re_{\tau} = 180$.}
    \label{tab:RMS_err_180}
\end{center}
\end{table}

\begin{figure}
\begin{center}
\includegraphics[width=.328\textwidth]{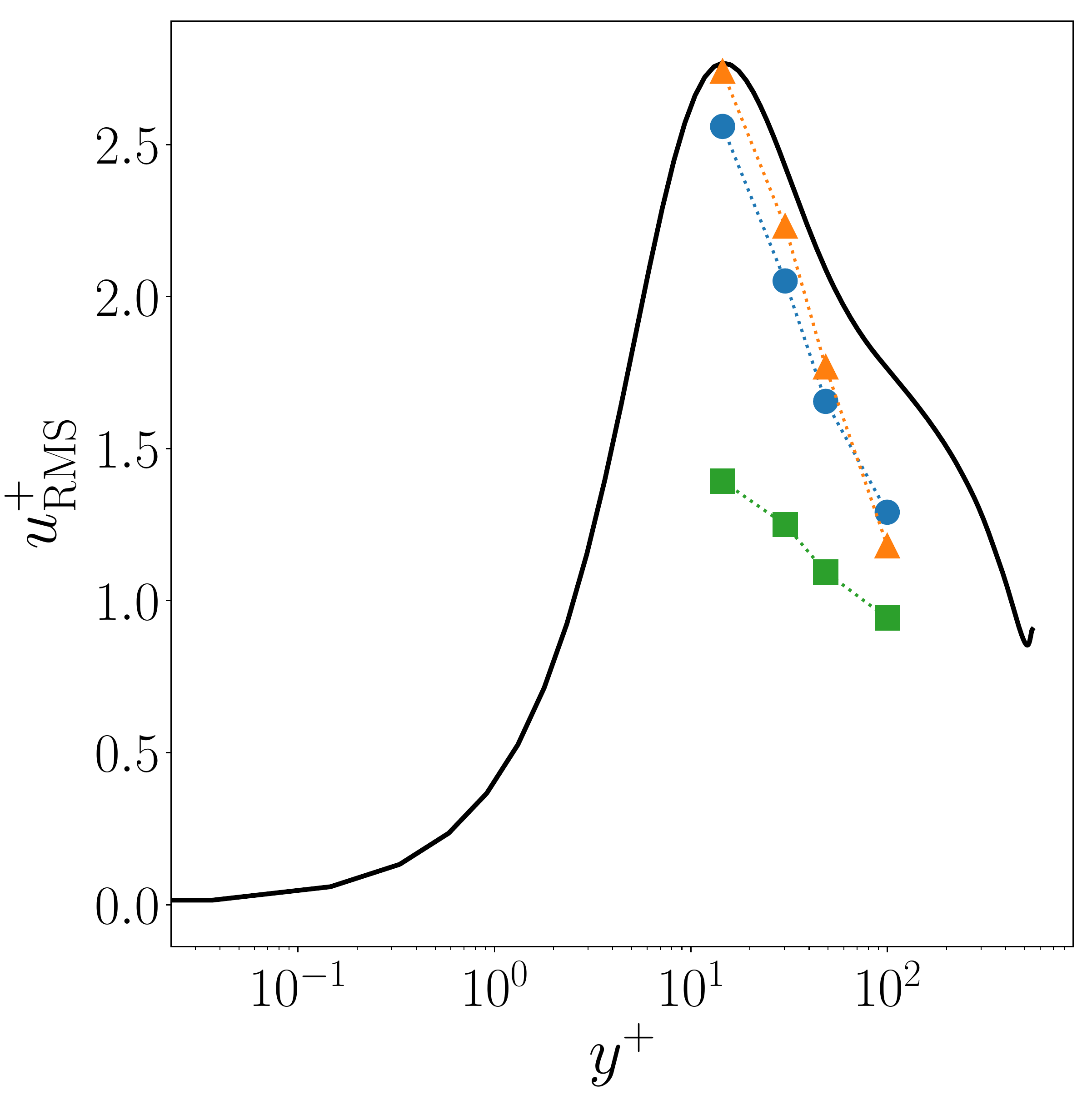}
\includegraphics[width=.328\textwidth]{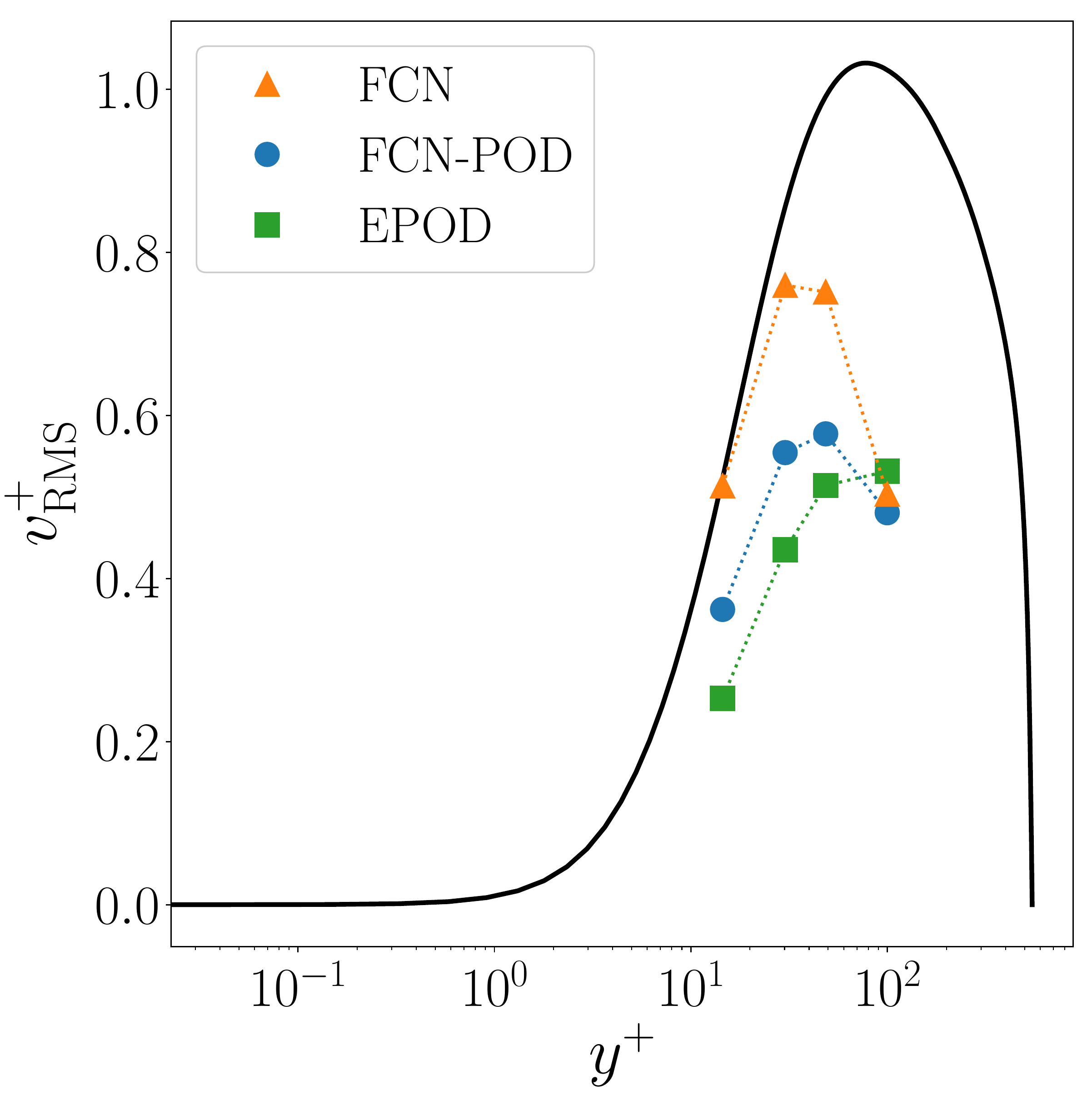}
\includegraphics[width=.328\textwidth]{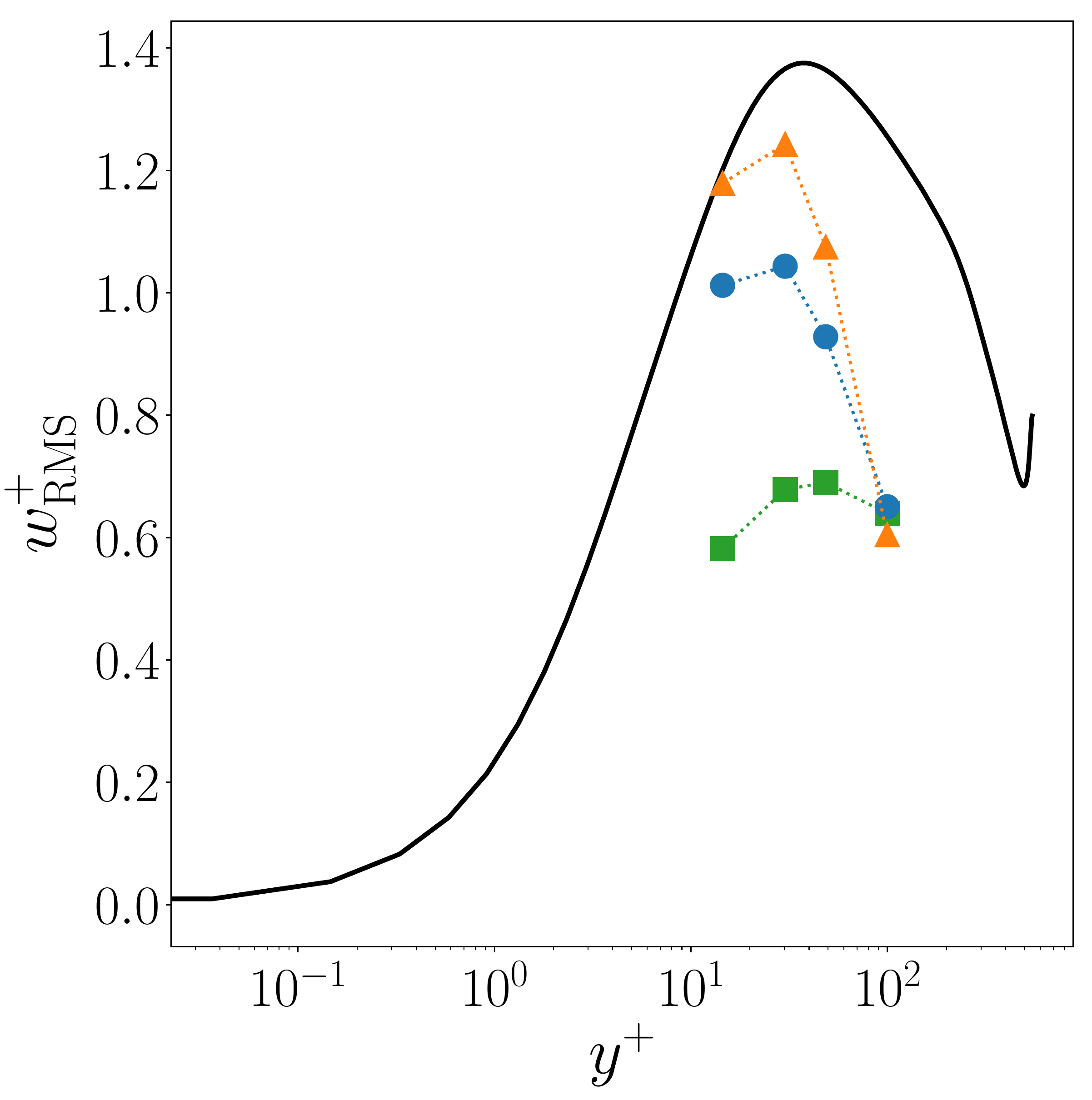}
\end{center}
\caption{\label{fig:stats550} Comparison between the DNS ( \full) and the predictions of streamwise (left),
 wall-normal (middle) and spanwise (right) velocity 
fluctuations at $Re_{\tau} = 550$.} 
\end{figure}

The statistical analysis is repeated also for the models trained at $Re_{\tau} = 550$, with the predicted RMS fluctuations shown in figure~\ref{fig:stats550} and the relative error with respect to the reference simulation in table~\ref{tab:RMS_err_550}. FCN-based models do not show a significant variation in the prediction of the streamwise fluctuations with respect to the results at $Re_{\tau}=180$, whereas the EPOD exhibits higher errors at this Reynolds number (also in the other two fluctuating components). The FCN has a consistent behaviour also for the fluctuations in the \textit{y}- and \textit{z}-directions, however the FCN-POD performs slightly worse than before, following the same trend but with higher error levels. The FCN-POD method outperforms the FCN approach only at $y^+=100$, confirming the results of the instantaneous performance at $Re_{\tau} = 550$. 

\begin{table}
\begin{center}
    \begin{tabular}{l l*{4}{c}}
        $E_\mathrm{RMS}^+(\cdot)\ [\%]$ &         & $y^+=15$                  & $y^+=30$                     & $y^+=50$                      & $y^+=100$          \\[0.3cm]
        
                                     & EPOD    & $49.69~~~~~~~~~~$            & $48.53~~~~~~~~~~~$           & $47.62~~~~~~~~~~~$            & $46.53~~~~~~~~~~~$ \\
        $u$                          & FCN     & $0.98~(\pm 0.66)$            & $\phantom{0}8.10~(\pm 0.62)$ & $15.33~(\pm 0.22)$            & $33.06~(\pm 0.30)$ \\
                                     & FCN-POD & $7.53~(\pm 0.27)$            & $15.53~(\pm 0.39)$           & $20.75~(\pm 0.85)$            & $26.79~(\pm 0.36)$ \\[0.3cm]
                                     
                                     & EPOD    & $51.57~~~~~~~~~~$            & $49.13~~~~~~~~~~~$           & $48.13~~~~~~~~~~~$            & $48.07~~~~~~~~~~~$ \\
        $v$                          & FCN     & $\phantom{0}1.74~(\pm 0.11)$ & $11.21~(\pm 1.41)$           & $24.20~(\pm 1.38)$            & $50.82~(\pm 0.26)$ \\
                                     & FCN-POD & $30.73~(\pm 0.06)$           & $35.20~(\pm 0.09)$           & $41.75~(\pm 0.10)$            & $53.04~(\pm 0.18)$ \\[0.3cm]
                                     
                                     & EPOD    & $51.56~~~~~~~~~~$            & $50.35~~~~~~~~~~~$           & $49.42~~~~~~~~~~~$            & $49.08~~~~~~~~~~~$ \\
        $w$                          & FCN     & $\phantom{0}1.86~(\pm 0.60)$ & $\phantom{0}9.03~(\pm 0.31)$ & $21.21~(\pm 1.27)$            & $51.83~(\pm 0.38)$ \\
                                     & FCN-POD & $15.74~(\pm 0.05)$           & $23.63~(\pm 0.07)$           & $31.97~(\pm 0.10)$            & $48.20~(\pm 0.24)$ \\                             
    \end{tabular}
    \caption{Percentage error in the prediction of the various RMS fluctations at the different wall-normal locations. Results at $Re_{\tau}=550$.}
    \label{tab:RMS_err_550}
\end{center}
\end{table}

\subsection{Predictions of power-spectral density}\label{ss:spec}
\begin{figure}
    \centerline{\includegraphics[width=384pt]{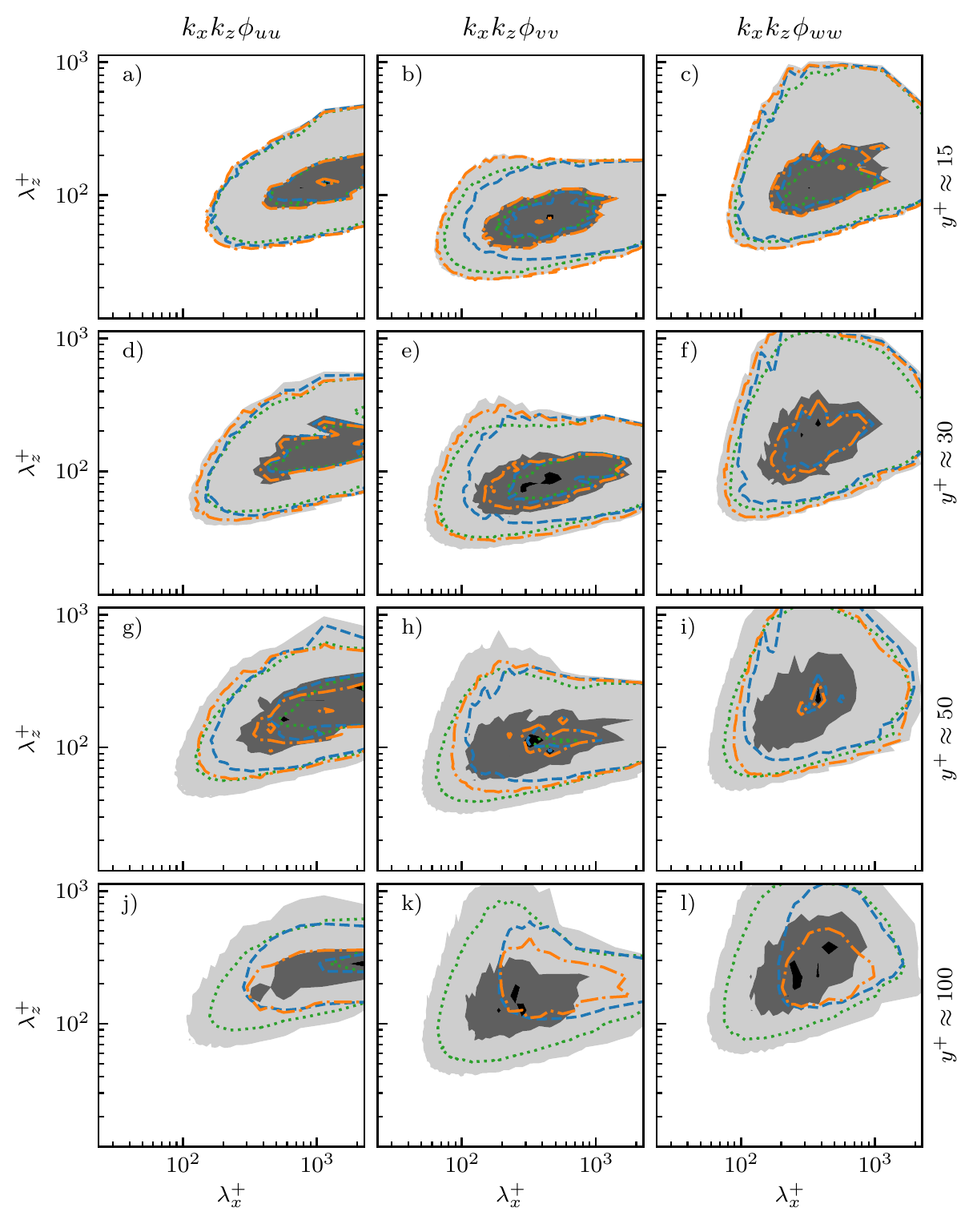}}
    \caption{Pre-multiplied two-dimensional power-spectral densities for $Re_{\tau}=180$. The contour levels contain 10\%, 50\% and 90\% of the maximum DNS power-spectral density. Shaded contours refer to the reference data, while contour lines refer to \lcap{.-}{fcn} FCN, \lcap{--}{FCN-POD} FCN-POD and \lcap{:}{epod} EPOD predictions, respectively.}
    \label{fig:spectra180}
\end{figure}

The energetic scales present in the predicted fields, as well as their associated energy, are compared with those in the reference DNS data through spectral analysis. In figure~\ref{fig:spectra180} we show the pre-multiplied two-dimensional power-spectral density of the streamwise, wall-normal and spanwise fluctuations (denoted by $\phi_{uu}$, $\phi_{vv}$ and $\phi_{ww}$ respectively) at $Re_{\tau}=180$, where $\lambda_{x}$ and $\lambda_{z}$ denote the streamwise and spanwise wavelengths, whereas $k_{x}$ and $k_{z}$ are the corresponding wavenumbers. These results confirm the observations made in $\S$\ref{ss:inst}: at $y^+=15$, all the considered models are able to correctly predict the energy content of the flow at all wavelengths, with the FCN slightly outperforming the two POD-based approaches. Note that the FCN-POD model is able to reconstruct the energy content of the flow at wavelengths that are longer than the size of the subdomains, proving that this is not a limiting factor for the model. However, a small jump, probably due to a lack of smoothness at the edges of the subdomains, can be observed in the streamwise wavelength for the 10\%-energy level in the wall-normal and spanwise components. These jumps are found at a wavelength $\lambda_x^+ \approx 180$, corresponding to the subdomain size.
At $y^+=30$ there is a slight energy attenuation which becomes increasingly more noticeable when the predicted flow is farther away from the wall. At $y^+=100$ the POD-based methods perform better than the FCN model, a fact that can be explained by considering two concurring aspects: the first is that POD methods only predict the temporal dynamics of the system, thus the overall energy-scale distribution stored in the POD spatial modes does not need to be predicted. This allows to reconstruct more than $50\%$ of the flow fields, at least in the streamwise component. The second aspect is the fact that the receptive field from the FCN method, while sufficient for planes closer to the wall, is not large enough to reproduce the large scales present at larger $y^{+}$.

\begin{figure}
    \centerline{\includegraphics[width=384pt]{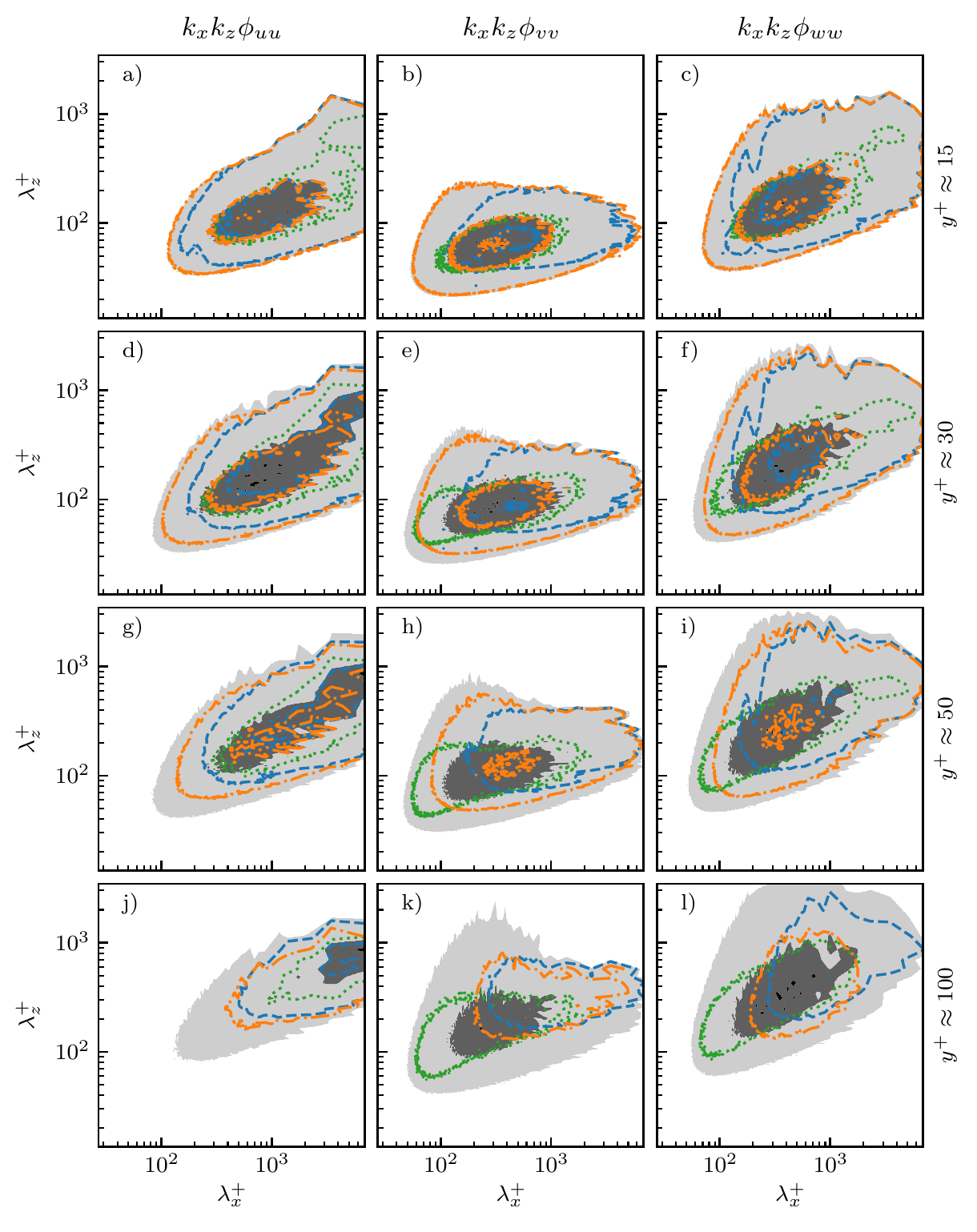}}
    \caption{Pre-multiplied two-dimensional power-spectral densities for $Re_{\tau}=550$. The contour levels contain 10\%, 50\% and 90\% of the maximum DNS power-spectral density. Shaded contours refer to the reference data, while contour lines refer to \lcap{.-}{fcn} FCN, \lcap{--}{FCN-POD} FCN-POD and \lcap{:}{epod} EPOD predictions, respectively.}
    \label{fig:spectra550}
\end{figure}

Figure~\ref{fig:spectra550} reports the reference and predicted power-spectral densities at $Re_{\tau}=550$. As opposed to what was observed for the instantaneous predictions and the turbulence statistics, the spectra highlight the differences and similarities between the models used for the two Reynolds numbers. As noted above, the FCN architecture is the same for both $Re_{\tau}$, however this implies that the receptive field is smaller at higher Reynolds number when it is measured in outer units. This can potentially help in the prediction of the small scales, but it can also be detrimental for the larger scales. Note that the different size of the receptive field does not seem to affect the predicted energy content of the FCN, which shows the same trends as for the low-$Re_{\tau}$ case. On the other hand, the spectra of the FCN-POD with $12\times12$ subdomains (the same amount as in the low-$Re_{\tau}$ case, not shown) exhibits spurious periodic peaks due to the tiling. This observation motivated the increased number of subdomains considered at $Re_{\tau}=550$.
As in the low-$Re_{\tau}$ case, the FCN-POD approach is able to reconstruct the scales larger than the subdomain size, but the countour line exhibits a small jump in the streamwise wavelength for the 10\%-energy level. This jump appears at a wavelength $\lambda_x^+ \approx 200$, corresponding to the subdomain size employed in the high-$Re_{\tau}$ case. This jump is also appreciated in the streamwise component at $y^+=15$.
The POD-based approaches are outperformed by the FCN in the range of $y^+=15-50$.
Farther from the wall, the accuracy of the FCN is matched by the FCN-POD method. It is interesting to note that the EPOD does not follow the same attenuation process as the FCN-based methods in the wall-normal and spanwise components. As one moves farther from the wall, the FCN-based methods fail to reproduce a wider range of small scales, whereas the EPOD exhibits more difficulties predicting the large scales. When it comes to the spectral peaks far from the wall, while the FCN-based methods produce noisier predictions than in the low-$Re_{\tau}$ case, the EPOD is not able to reproduce that part of the spectra.

\section{Transfer learning}\label{ss:tl}

A number of more advanced techniques have also started to be adopted from the specialized machine-learning literature and applied to fluid-dynamics research. One notable example is \textit{transfer learning}~\citep{pan2009survey}, a method that allows to transfer knowledge from one neural-network model to another one, thus reducing the amount data and time required for training. \citet{guastoni} showed that the training time at a given wall-normal location may be significantly reduced if the network parameters are initialized using the optimized parameters of a previously-trained network at another wall-normal location. Similarly, \citet{kim2020prediction} used the convolutional network trained at a low Reynolds number to predict the flow at a higher Reynolds number.

Transfer learning represents an appealing solution for the main drawback of neural networks, which is the need to train them with a sufficient amount of data. Training typically requires specialized hardware and in our specific application the computational cost of generating the training and test datasets is not negligible. Furthermore, this cost grows as $Re_{\tau}$ increases, making the generation of training data through DNS unfeasible at the Reynolds numbers that are relevant for engineering applications. In this regard, it is important to make an efficient use of the data and the trained models at our disposal. 
In this work, the possibility of transferring knowledge between models trained at different friction Reynolds numbers is investigated. At a fixed wall-normal distance, the weights of the FCN model trained on the dataset at $Re_{\tau} = 180$ are loaded before training the network with the higher-$Re_{\tau}$ dataset. This is possible because the network has the same amount of trainable parameters in both cases, as noted above. The learning rate is the only parameter that needs to be modified: a lower value has to be set, in order to prevent the optimizer from diverging too quickly from the weight configuration used for initialization. While in \citet{guastoni} we froze the first layers of the initialized network because the input was the same at the different wall-normal locations, in this case all the layers are trainable because the input distribution changes from one $Re_{\tau}$ to the other one. 

\begin{figure}
    \centerline{\includegraphics[width=\textwidth]{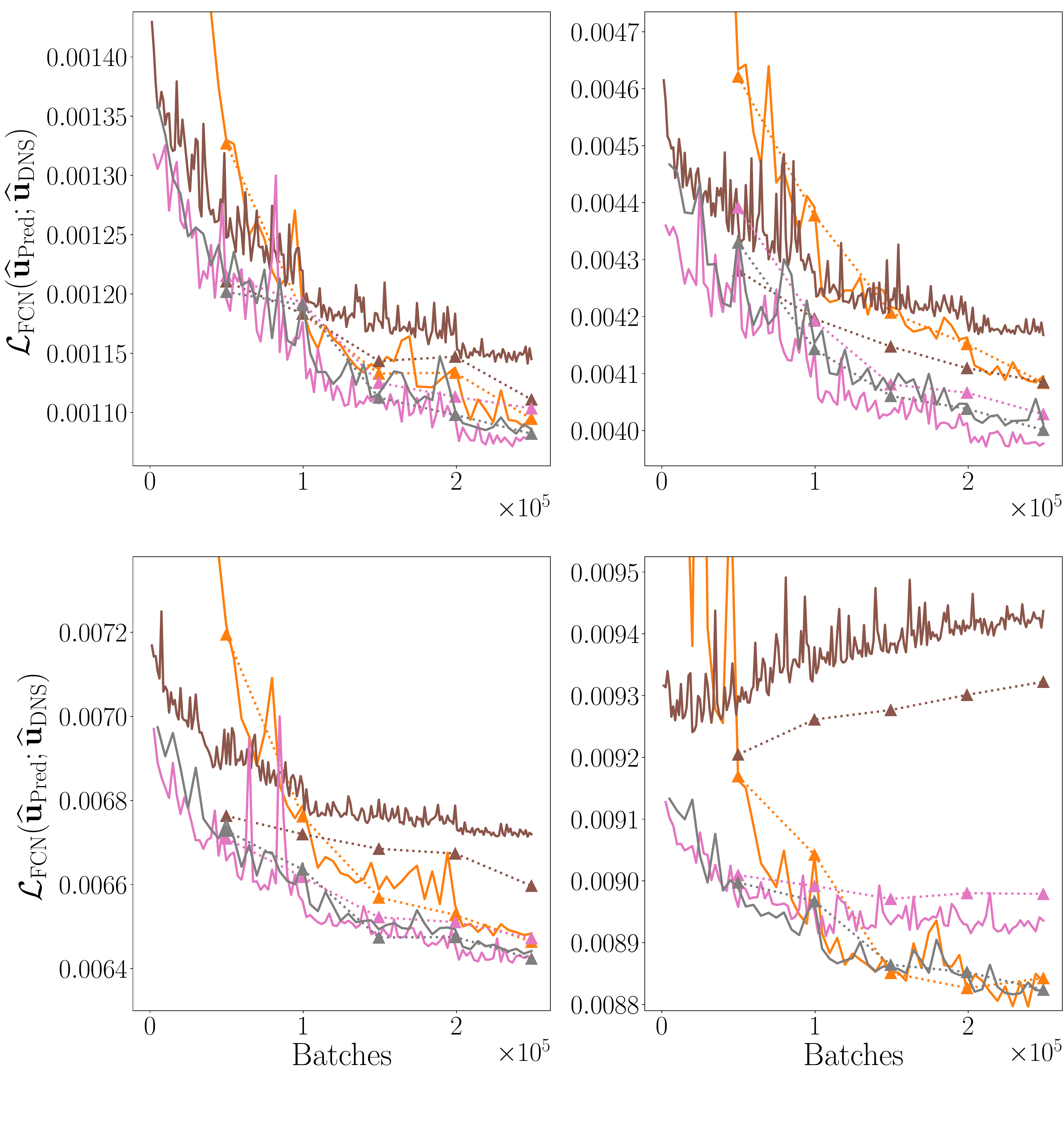}}
    \caption{Validation (\full) and test (\dashed) loss in the FCN prediction at (from left to right, top to bottom): $y^+=15$, 30, 50 and 100. Orange represents the models trained with the full dataset and random initialization, grey the models trained with the full dataset and initialized with previously-trained networks, pink and brown represent models initialized with the parameters from the $Re_{\tau}=180$ network, trained with 50\% and 25\% of the original dataset, respectively.} 
    \label{fig:transfer}
\end{figure}

First we considered an initialized model with the full training/validation dataset, in order to assess the effect of the initialization on the training results. Subsequently, new training runs of the initialized model were performed  with 25\% and 50\% of the original dataset. Differently from the previous sections, only one training run was performed for each case. In order to compare models trained with datasets of different sizes, we considered the number of weight updates through the optimization algorithm during training. In figure~\ref{fig:transfer} the validation and test losses are compared for the models trained with the full dataset and a random initialization. 
When the initialized model is trained on the full dataset, the performance is consistently better than that of the random initialization, both in terms of validation and test loss. The improvement is more evident close to the wall, whereas at $y^+=100$ the two models provide approximately the same results after the first 150,000 updates. Transferring knowledge between different Reynolds numbers is then not only feasible, but also advantageous in terms of performance when the same amount of data is considered.

If the training/validation dataset is reduced, the validation loss seems to overestimate the error compared with that of the test dataset when 25\% of the data is used, at all wall-normal locations. The opposite holds when 50\% of the training dataset is used, instead. Up to $y^+=50$, the initialized networks are able to provide a performance that is very similar to that of the reference model with the same number of updates, with significant savings in terms of amount of data needed to train the network. On the other hand, at $y^+=100$ a sufficient number of samples becomes a necessary condition to ensure the convergence to an optimal configuration: the loss of the network trained with 50\% of the training dataset does not improve after the first 100,000 updates, while with 25\% of the original dataset the network exhibits overfitting.

The initialized models are able to provide a comparable accuracy also from the statistical point of view, as reported in table~\ref{tab:RMS_err_transfer}. We stress once again that the networks are not explicitly optimized to reduce the error in the statistics and that small variations in these error figures can be ascribed to the stochastic nature of the optimization algorithm. Overall, these results demonstrate the feasibility of knowledge transfer from models at different $Re_{\tau}$: with careful tuning of the hyperparameters it should be possible to substantially reduce the training time, as well as the amount of data needed for training. Although not tested, the transfer between different wall-normal locations described in~\citet{guastoni} is still applicable in this case, thus enabling a more efficient prediction of the flow at different wall-normal locations. 

\begin{table}
\begin{center}
    \begin{tabular}{l l*{4}{c}}
        $E_\mathrm{RMS}^+(\cdot)\ [\%]$ &          & $y^+=15$                    & $y^+=30$                     & $y^+=50$                      & $y^+=100$                    \\[0.3cm]
        
        $u$                             & Ref.     & $0.98~(\pm 0.66)$           & $8.10~(\pm 0.62)$            & $15.33~(\pm 0.22)$            & $33.06~(\pm 0.30)$           \\
                                        & 100\%    & $1.22\phantom{~(\pm 0.00)}$ & $7.15\phantom{~(\pm 0.00)}$  & $16.09\phantom{~(\pm 0.00)}$  & $32.75\phantom{~(\pm 0.00)}$ \\
                                        & 50\%     & $2.94\phantom{~(\pm 0.00)}$ & $7.11\phantom{~(\pm 0.00)}$  & $16.33\phantom{~(\pm 0.00)}$  & $34.11\phantom{~(\pm 0.00)}$ \\
                                        & 25\%     & $1.15\phantom{~(\pm 0.00)}$ & $7.74\phantom{~(\pm 0.00)}$  & $14.78\phantom{~(\pm 0.00)}$  & $33.78\phantom{~(\pm 0.00)}$ \\[0.3cm]
                                     
        $v$                             & Ref.     & $1.74~(\pm 0.11)$           & $11.21~(\pm 1.41)$           & $24.20~(\pm 1.38)$            & $50.82~(\pm 0.26)$           \\
                                        & 100\%    & $1.86\phantom{~(\pm 0.00)}$ & $9.40\phantom{0~(\pm 0.00)}$ & $24.40\phantom{~(\pm 0.00)}$  & $51.59\phantom{~(\pm 0.00)}$ \\                            
                                        & 50\%     & $2.40\phantom{~(\pm 0.00)}$ & $9.46\phantom{0~(\pm 0.00)}$ & $25.96\phantom{~(\pm 0.00)}$  & $50.90\phantom{~(\pm 0.00)}$ \\
                                        & 25\%     & $1.71\phantom{~(\pm 0.00)}$ & $11.33\phantom{~(\pm 0.00)}$ & $23.15\phantom{~(\pm 0.00)}$  & $50.43\phantom{~(\pm 0.00)}$ \\[0.3cm]
                                     
        $w$                             & Ref.     & $1.86~(\pm 0.60)$           & $9.03~(\pm 0.31)$            & $21.21~(\pm 1.27)$            & $51.83~(\pm 0.38)$           \\
                                        & 100\%    & $1.75\phantom{~(\pm 0.00)}$ & $8.65\phantom{~(\pm 0.00)}$  & $21.05\phantom{~(\pm 0.00)}$  & $53.04\phantom{~(\pm 0.00)}$ \\
                                        & 50\%     & $2.61\phantom{~(\pm 0.00)}$ & $7.70\phantom{~(\pm 0.00)}$  & $21.34\phantom{~(\pm 0.00)}$  & $52.60\phantom{~(\pm 0.00)}$ \\
                                        & 25\%     & $1.22\phantom{~(\pm 0.00)}$ & $9.67\phantom{~(\pm 0.00)}$  & $20.35\phantom{~(\pm 0.00)}$  & $49.75\phantom{~(\pm 0.00)}$ \\                             
    \end{tabular}
    \caption{Percentage error in the prediction of the various RMS fluctations at the different wall-normal locations from models initialized with parameters from the $Re_{\tau}=180$ FCN. The statistics of the different initialized models are computed after 250,000 updates, and they are shown together the models with full dataset and random initialization, which is included as a reference. Results at $Re_{\tau}=550$.}
    \label{tab:RMS_err_transfer}
\end{center}
\end{table}

\section{Conclusions}\label{ss:conclu}

In this work, we introduced and compared two different models based on fully-convolutional neural networks, for prediction of the velocity fluctuations at a given wall-normal distance, using quantities measured at the wall as inputs. The FCN and FCN-POD models are improved versions of previous architectures, used by \citet{guastoni} and \citet{Guemes2019sensing}, respectively. Both of them are able to provide predictions in very good agreement with the reference data, simulated by means of the pseudo-spectral DNS code SIMSON~\citep{chevalier}, up to $y^{+}=50$. Such an agreement is verified by comparing the error in instantaneous predictions, turbulence statistics (namely RMS fluctuations) and the energy content at the different wavelengths ({\it i.e.} spectral analysis). Both models show better prediction capabilities than EPOD (which is a linear method) in almost all the wall-normal locations and investigated features, thanks to their ability to predict nonlinear scale interactions. 
Furthermore, we showed that these architectures can be used at two different friction Reynolds numbers ($Re_{\tau}=180$ and 550) with minimal modifications, providing satisfactory results on both datasets.

The two models are designed under the assumption that local information at the wall is sufficient to predict the flow farther away, however the FCN-POD model partially encodes further physical information of the system by using the spatial modes obtained through POD of the training dataset. On the other hand, features like the periodicity of the flow are enforced in the FCN by exploiting the mathematical characteristics of the model. 
These architectural differences are associated with performance discrepancies at the tested wall-normal locations: the FCN provides higher accuracy than the FCN-POD model closer to the wall, {\it i.e.} up to $y^+=30$ at $Re_{\tau} = 180$ and up to $y^+=50$ at $Re_{\tau} = 550$. Farther from the wall, the FCN-POD method produces the most accurate predictions. The choice between these two models is motivated by the application into which the prediction model is integrated.

Despite the encouraging results discussed here, both models can be improved in terms of network architecture and training. An attempt to embed further physical information into the FCN did not result in improved predictions, as reported in $\S$\ref{sss:shift}. The correct way of incorporating this information to enhance the predictions is an active area of research. As another example, the high-frequency noise in the FCN predictions could be reduced with appropriate filtering, possibly adding a trainable layer to the network to perform this operation. The FCN-POD model has a higher number of hyperparameters to be set, such as the number of predicted temporal modes or the size of the subdomains. A more thorough inspection of the hyperparameter space may provide a significant improvement in the prediction performance. Differently from the FCN, in the FCN-POD model the velocity components are not scaled to have the same magnitude: such a modification could help to predict the wall-normal and spanwise components of the velocity more accurately, even though it would also modify the POD mode sorting because of the different energy norm. Furthermore, the FCN-POD results exhibit lack of smoothness at the subdomain edges in the flow predictions. Finally, both models are trained to minimize a loss function based on the instantaneous error. Such a function could be modified to improve other physical characteristics of the predicted flow, for example the turbulence statistics and the spectral energy content.

To reduce the training time in view of industrial applications, the implementation of transfer learning was tested for the FCN model. Transfer learning can exploit a network trained at a lower Reynolds number to provide the weight initialization for training at a higher Reynolds number, thus reducing the requirements in terms of training time and data. The results are very encouraging, showing that it is possible to train the network with 50\% and even 25\% of the original training dataset, obtaining a performance similar to that of the reference model up to $y^{+}=50$.

Once the neural networks are trained, they are computationally cheap to evaluate, and they can become even cheaper by \textit{pruning} the parts that have a negligible contribution to the final result off the network. Such an operation is not possible \textit{a priori}, since the training determines how the inputs have to be processed to obtain the output. By reducing the computational cost of the evaluation it is possible to deploy the model using low-powered hardware and/or potentially run it in real-time. Thus, the proposed FCN-based methods could be used for non-intrusive sensing of the flow, which is needed for closed-loop control applications.  Furthermore, since the FCN models are able to reproduce non-linear interactions in wall-bounded turbulence, new promising avenues in turbulence research could be opened by the network interpretation~\citep{fan2020interpretability}, as shown by \citet{iten2020discovering}, who demonstrated that neural networks can provide relevant physical insights.

\section*{Acknowledgements}
All the codes employed in this work will be released as open-source on GitHub upon publication in the peer-reviewed literature. The authors acknowledge the funding provided by the Swedish e-Science Research Centre (SeRC), the G\"oran Gustafsson Foundation and the Knut and Alice Wallenberg (KAW) Foundation. The numerical simulations were carried out on resources provided by the Swedish National Infrastructure for Computing (SNIC) at PDC and HPC2N. Part of study was conducted in the context of the 4th Madrid Turbulence Workshop, funded by the COTURB project (Coherent Structures in Wall-bounded Turbulence), funded by the European Research Council (ERC), under
grant ERC-2014.AdG-669505. SD and AI were partially supported by the Grant DPI2016-79401-R funded by the Spanish State Research Agency (SRA) and European Regional Development Fund (ERDF).

\section*{Declaration of Interests}
The authors report no conflict of interest.

\appendix
\section{}\label{appA}
This Appendix contains the wall-normal and spanwise fluctuations corresponding to the fields shown in figures~\ref{fig:field_comp180} and~\ref{fig:field_comp550} for $Re_{\tau}=180$ and 550, respectively.

\begin{figure}
\begin{center}
\includegraphics[width=\textwidth]{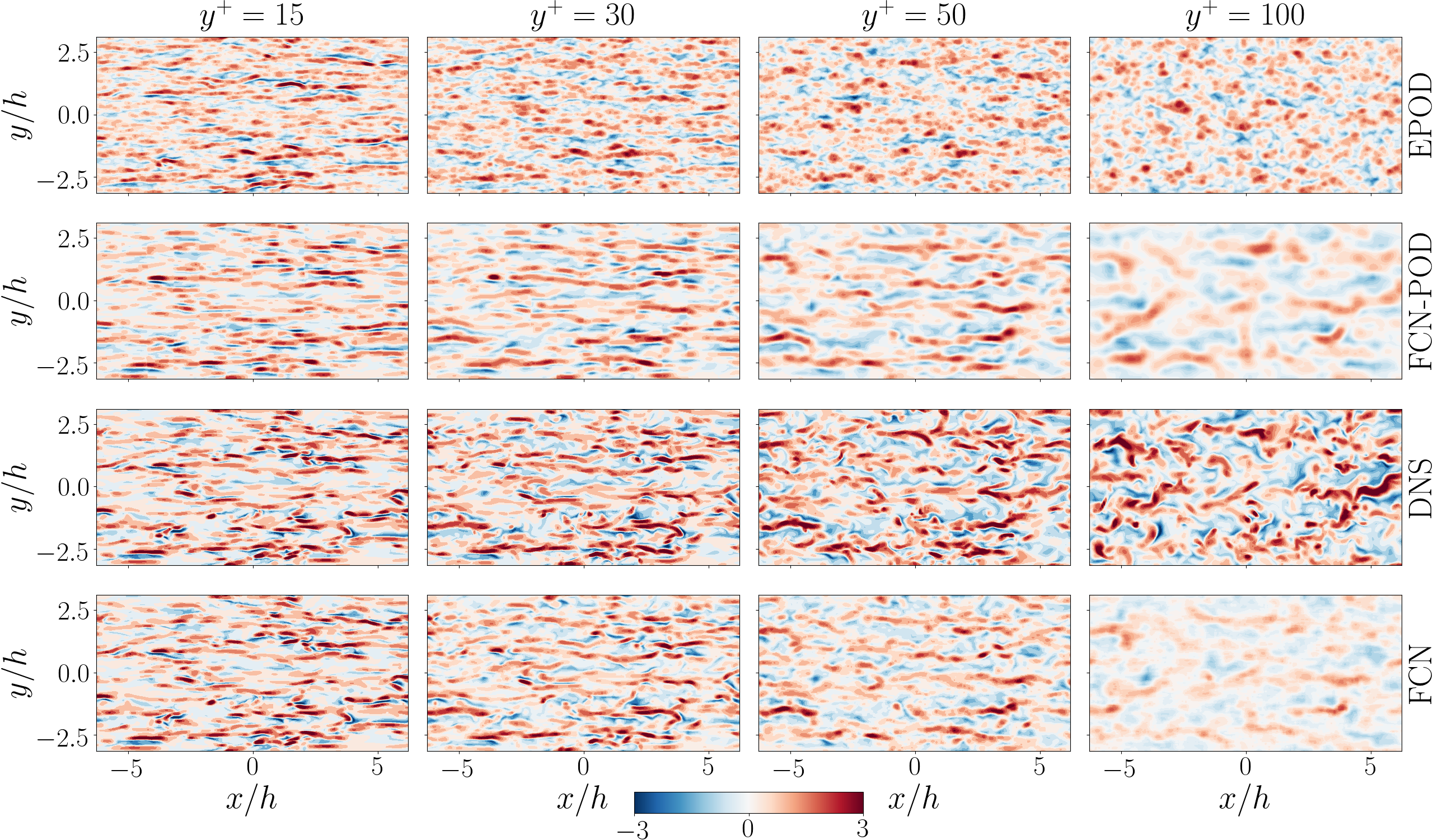}
\end{center}
\caption{\label{fig:field_comp180v} Comparison of the wall-normal fluctuation fields at $Re_{\tau} = 180$, scaled with the corresponding $v_\mathrm{RMS}$, from EPOD (1$^{\text{st}}$ row), FCN-POD (2$^{\text{nd}}$ row), reference DNS (3$^{\text{rd}}$ row) and FCN (4$^{\text{th}}$ row). Results at $y^+=15$ (1$^{\text{st}}$ column), $y^+=30$ (2$^{\text{nd}}$ column), $y^+=50$ (3$^{\text{rd}}$ column) and $y^+=100$ (4$^{\text{th}}$ column).}
\end{figure}

\begin{figure}
\begin{center}
\includegraphics[width=\textwidth]{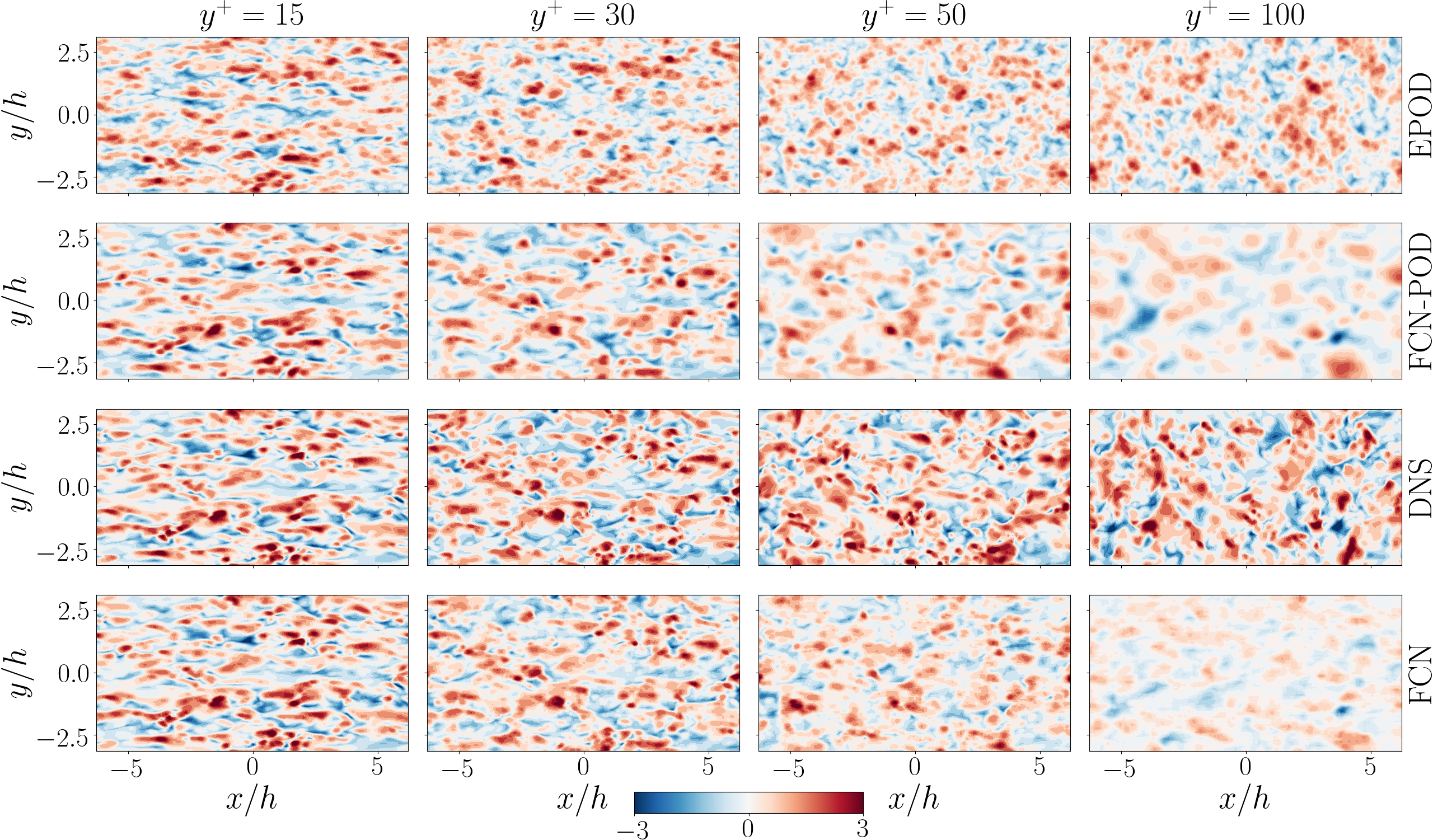}
\end{center}
\caption{\label{fig:field_comp180w} Comparison of the spanwise fluctuation  fields at $Re_{\tau} = 180$, scaled with the corresponding $w_\mathrm{RMS}$, from EPOD (1$^{\text{st}}$ row), FCN-POD (2$^{\text{nd}}$ row), reference DNS (3$^{\text{rd}}$ row) and FCN (4$^{\text{th}}$ row). Results at $y^+=15$ (1$^{\text{st}}$ column), $y^+=30$ (2$^{\text{nd}}$ column), $y^+=50$ (3$^{\text{rd}}$ column) and $y^+=100$ (4$^{\text{th}}$ column).}
\end{figure}

\begin{figure}
\begin{center}
\includegraphics[width=\textwidth]{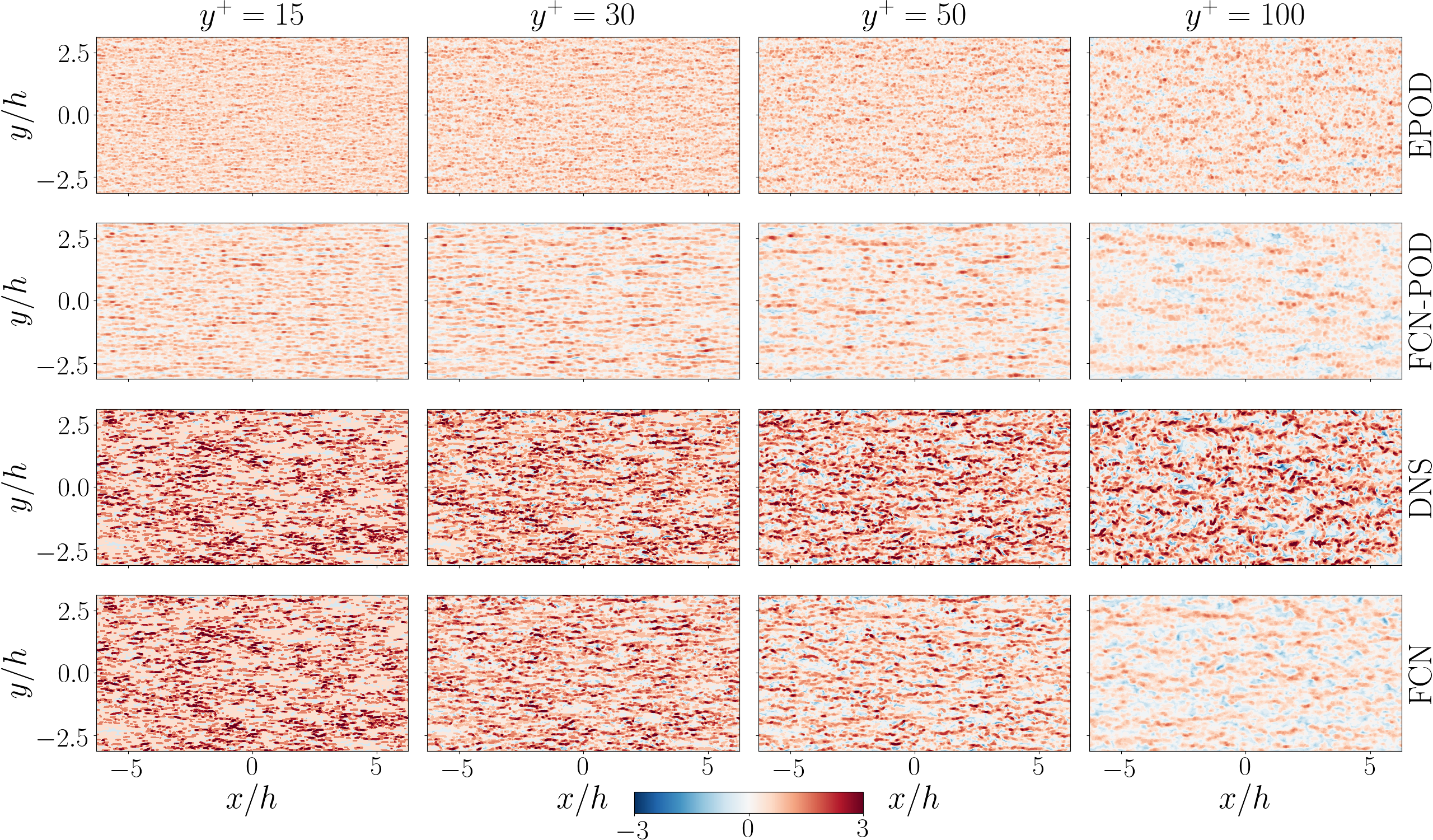}
\end{center}
\caption{\label{fig:field_comp550v} Comparison of the wall-normal fluctuation fields at $Re_{\tau} = 550$, scaled with the corresponding $v_\mathrm{RMS}$, from EPOD (1$^{\text{st}}$ row), FCN-POD (2$^{\text{nd}}$ row), reference DNS (3$^{\text{rd}}$ row) and FCN (4$^{\text{th}}$ row). Results at $y^+=15$ (1$^{\text{st}}$ column), $y^+=30$ (2$^{\text{nd}}$ column), $y^+=50$ (3$^{\text{rd}}$ column) and $y^+=100$ (4$^{\text{th}}$ column).}
\end{figure}

\begin{figure}
\begin{center}
 \includegraphics[width=\textwidth]{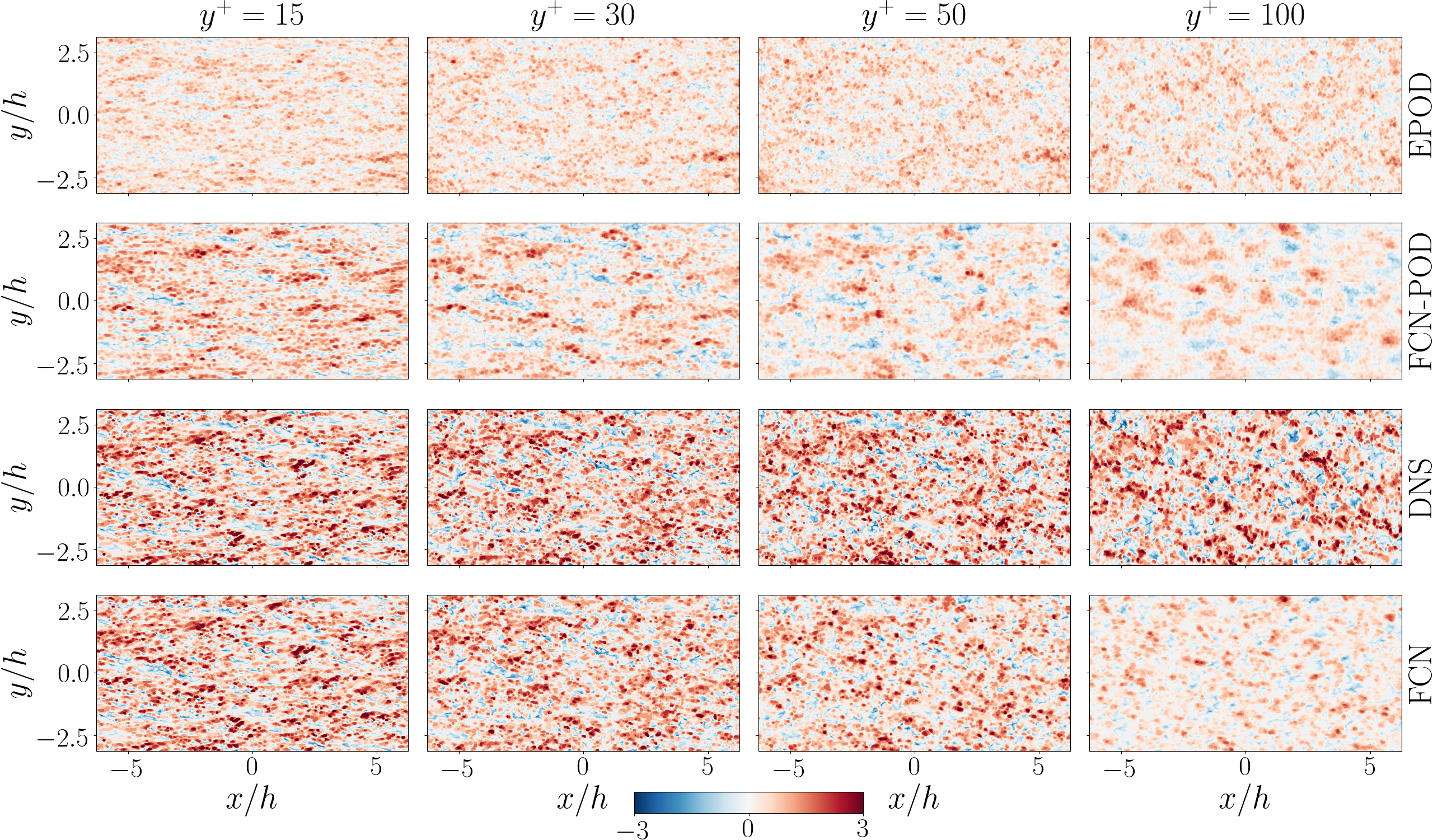}
 \end{center}
\caption{\label{fig:field_comp550w} Comparison of the spanwise fluctuation  fields at $Re_{\tau} = 550$, scaled with the corresponding $w_\mathrm{RMS}$, from EPOD (1$^{\text{st}}$ row), FCN-POD (2$^{\text{nd}}$ row), reference DNS (3$^{\text{rd}}$ row) and FCN (4$^{\text{th}}$ row). Results at $y^+=15$ (1$^{\text{st}}$ column), $y^+=30$ (2$^{\text{nd}}$ column), $y^+=50$ (3$^{\text{rd}}$ column) and $y^+=100$ (4$^{\text{th}}$ column).}
\end{figure}

\bibliographystyle{jfm}
\bibliography{bibliography}

\begin{thebibliography}{79}
\expandafter\ifx\csname natexlab\endcsname\relax\def\natexlab#1{#1}\fi
\def\au#1{#1} \def\ed#1{#1} \def\yr#1{#1}\def\at#1{#1}\def\jt#1{\textit{#1}}
  \def\bt#1{#1}\def\bvol#1{\textbf{#1}} \def\vol#1{#1} \def\pg#1{#1}
  \def\publ#1{#1}\def\arxiv#1{#1}\def\org#1{#1}\def\st#1{\textit{#1}}

\bibitem[Abadi {\em et~al.\/}(2016)Abadi, Barham, Chen, Chen, Davis, Dean,
  Devin, Ghemawat, Irving, Isard {\em et~al.\/}]{abadi2016tensorflow}
{\sc \au{Abadi, M.}, \au{Barham, P.}, \au{Chen, J.}, \au{Chen, Z.}, \au{Davis,
  A.}, \au{Dean, J.}, \au{Devin, M.}, \au{Ghemawat, S.}, \au{Irving, G.},
  \au{Isard, M.} \& \au{others}} \yr{2016} Tensorflow: A system for large-scale
  machine learning.  \bt{In {\em 12th {USENIX} Symposium on Operating Systems
  Design and Implementation ({OSDI} 16)\/}},  \pg{pp. 265--283}.

\bibitem[Adrian(1988)]{adrian88}
{\sc \au{Adrian, R.}} \yr{1988}  \at{Stochastic estimation of organized
  turbulent structure: Homogeneous shear flow}.  \jt{J. Fluid Mech.}
  \bvol{190},  \pg{531--559}.

\bibitem[Baars \& Tinney(2014)]{baars2014proper}
{\sc \au{Baars, W.~J.} \& \au{Tinney, C.~E.}} \yr{2014}  \at{Proper orthogonal
  decomposition-based spectral higher-order stochastic estimation}.  \jt{Phys.
  Fluids}  \bvol{26}~(5),  \pg{055112}.

\bibitem[Baldi \& Hornik(1989)]{baldi1989neural}
{\sc \au{Baldi, P.} \& \au{Hornik, K.}} \yr{1989}  \at{Neural networks and
  principal component analysis: Learning from examples without local minima}.
  \jt{Neural Networks}  \bvol{2}~(1),  \pg{53--58}.

\bibitem[Beck {\em et~al.\/}(2019)Beck, Flad \& Munz]{beck_et_al}
{\sc \au{Beck, A.~D.}, \au{Flad, D.~G.} \& \au{Munz, C.-D.}} \yr{2019}
  \at{{Deep neural networks for data-driven LES closure models}}.  \jt{J.
  Comput. Phys.}  \bvol{398},  \pg{108910}.

\bibitem[Bor{\'e}e(2003)]{boree2003extended}
{\sc \au{Bor{\'e}e, J.}} \yr{2003}  \at{Extended proper orthogonal
  decomposition: a tool to analyse correlated events in turbulent flows}.
  \jt{Exp. Fluids}  \bvol{35}~(2),  \pg{188--192}.

\bibitem[Bourgeois {\em et~al.\/}(2013)Bourgeois, Noack \&
  Martinuzzi]{bourgeois2013generalized}
{\sc \au{Bourgeois, JA}, \au{Noack, BR} \& \au{Martinuzzi, RJ}} \yr{2013}
  \at{Generalized phase average with applications to sensor-based flow
  estimation of the wall-mounted square cylinder wake}.  \jt{J. Fluid Mech.}
  \bvol{736},  \pg{316--350}.

\bibitem[Bourlard \& Kamp(1988)]{bourlard1988auto}
{\sc \au{Bourlard, H.} \& \au{Kamp, Y.}} \yr{1988}  \at{Auto-association by
  multilayer perceptrons and singular value decomposition}.  \jt{Biol. Cybern.}
   \bvol{59}~(4-5),  \pg{291--294}.

\bibitem[Brunton {\em et~al.\/}(2020)Brunton, Noack \&
  Koumoutsakos]{brunton2020machine}
{\sc \au{Brunton, S.~L.}, \au{Noack, B.~R.} \& \au{Koumoutsakos, P.}} \yr{2020}
   \at{Machine learning for fluid mechanics}.  \jt{Annu. Rev. Fluid Mech.}
  \bvol{52},  \pg{477--508}.

\bibitem[Bucci {\em et~al.\/}(2019)Bucci, Semeraro, Allauzen, Wisniewski,
  Cordier \& Mathelin]{bucci_et_al}
{\sc \au{Bucci, M.~A.}, \au{Semeraro, O.}, \au{Allauzen, A.}, \au{Wisniewski,
  G.}, \au{Cordier, L.} \& \au{Mathelin, L.}} \yr{2019}  \at{{Control of
  chaotic systems by deep reinforcement learning}}.  \jt{Proc. R. Soc. A}
  \bvol{475},  \pg{20190351}.

\bibitem[Chevalier {\em et~al.\/}(2007)Chevalier, Schlatter, Lundbladh \&
  Henningson]{chevalier}
{\sc \au{Chevalier, M.}, \au{Schlatter, P.}, \au{Lundbladh, A.} \&
  \au{Henningson, D.~S.}} \yr{2007}  \bt{A pseudospectral solver for
  incompressible boundary layer flows}. {\em Tech. Rep.\/}.  \org{TRITA-MEK
  2007:07. KTH Mechanics, Stockholm, Sweden}.

\bibitem[Choi {\em et~al.\/}(1994)Choi, Moin \& Kim]{choi_et_al}
{\sc \au{Choi, H.}, \au{Moin, P.} \& \au{Kim, J.}} \yr{1994}  \at{Active
  turbulence control for drag reduction in wall-bounded flows}.  \jt{J. Fluid
  Mech.}  \bvol{262},  \pg{75--110}.

\bibitem[De~Fauw {\em et~al.\/}(2018)De~Fauw, Ledsam \&
  Romera-Paredes]{defauw_et_al_2018}
{\sc \au{De~Fauw, J.}, \au{Ledsam, J.R.} \& \au{Romera-Paredes, B. et~al.}}
  \yr{2018}  \at{{Clinically applicable deep learning for diagnosis and
  referral in retinal disease}}.  \jt{Nat. Med.}  \bvol{24},  \pg{1342--1350}.

\bibitem[Discetti {\em et~al.\/}(2019)Discetti, Bellani, {\"O}rl{\"u},
  Serpieri, Sanmiguel~Vila, Raiola, Zheng, Mascotelli, Talamelli \&
  Ianiro]{discetti2019characterization}
{\sc \au{Discetti, S.}, \au{Bellani, G.}, \au{{\"O}rl{\"u}, R.}, \au{Serpieri,
  J.}, \au{Sanmiguel~Vila, C.}, \au{Raiola, M.}, \au{Zheng, X.},
  \au{Mascotelli, L.}, \au{Talamelli, A.} \& \au{Ianiro, A.}} \yr{2019}
  \at{Characterization of very-large-scale motions in high-re pipe flows}.
  \jt{Exp. Therm. Fluid Sci.}  \bvol{104},  \pg{1--8}.

\bibitem[Discetti {\em et~al.\/}(2018)Discetti, Raiola \&
  Ianiro]{discetti2018estimation}
{\sc \au{Discetti, S.}, \au{Raiola, M.} \& \au{Ianiro, A.}} \yr{2018}
  \at{Estimation of time-resolved turbulent fields through correlation of
  non-time-resolved field measurements and time-resolved point measurements}.
  \jt{Exp. Therm. Fluid Sci.}  \bvol{93},  \pg{119--130}.

\bibitem[Dogan {\em et~al.\/}(2019)Dogan, \"Orl\"u, Gatti, Vinuesa \&
  Schlatter]{dogan_et_al}
{\sc \au{Dogan, E.}, \au{\"Orl\"u, R.}, \au{Gatti, D.}, \au{Vinuesa, R.} \&
  \au{Schlatter, P.}} \yr{2019}  \at{Quantification of amplitude modulation in
  wall-bounded turbulence}.  \jt{Fluid Dyn. Res.}  \bvol{51},  \pg{011408}.

\bibitem[Dumoulin \& Visin(2016)]{dumoulin2016guide}
{\sc \au{Dumoulin, V.} \& \au{Visin, F.}} \yr{2016}  \at{A guide to convolution
  arithmetic for deep learning}.  \jt{arXiv preprint arXiv:1603.07285} .

\bibitem[Duraisamy {\em et~al.\/}(2019)Duraisamy, Iaccarino \&
  Xiao]{duraisamy_et_al}
{\sc \au{Duraisamy, K.}, \au{Iaccarino, G.} \& \au{Xiao, H.}} \yr{2019}
  \at{Turbulence modeling in the age of data}.  \jt{Annu. Rev. Fluid Mech.}
  \bvol{51},  \pg{357--377}.

\bibitem[Eivazi {\em et~al.\/}(2020)Eivazi, Guastoni, Schlatter, Azizpour \&
  Vinuesa]{eivazi2020recurrent}
{\sc \au{Eivazi, H.}, \au{Guastoni, L.}, \au{Schlatter, P.}, \au{Azizpour, H.}
  \& \au{Vinuesa, R.}} \yr{2020}  \at{{Recurrent neural networks and
  Koopman-based frameworks for temporal predictions in turbulence}}.  \jt{arXiv
  preprint arXiv:2005.02762} .

\bibitem[Encinar \& Jim\'enez(2019)]{encinar}
{\sc \au{Encinar, M.~P.} \& \au{Jim\'enez, J.}} \yr{2019}
  \at{Logarithmic-layer turbulence: {A} view from the wall}.  \jt{Phys. Rev.
  Fluids}  \bvol{4},  \pg{114603}.

\bibitem[Erichson {\em et~al.\/}(2020)Erichson, Mathelin, Yao, Brunton, Mahoney
  \& Kutz]{erichson_et_al}
{\sc \au{Erichson, N.~B.}, \au{Mathelin, L.}, \au{Yao, Z.}, \au{Brunton,
  S.~L.}, \au{Mahoney, M.~W.} \& \au{Kutz, J.~N.}} \yr{2020}  \at{Shallow
  neural networks for fluid flow reconstruction with limited sensors}.
  \jt{Proc. R. Soc. A}  \bvol{476},  \pg{20200097}.

\bibitem[Fan {\em et~al.\/}(2020)Fan, Xiong \& Wang]{fan2020interpretability}
{\sc \au{Fan, F.}, \au{Xiong, J.} \& \au{Wang, G.}} \yr{2020}  \at{On
  interpretability of artificial neural networks}.  \jt{arXiv preprint
  arXiv:2001.02522} .

\bibitem[Fukami {\em et~al.\/}(2019{\natexlab{{\em a\/}}})Fukami, Fukagata \&
  Taira]{fukami2019super}
{\sc \au{Fukami, K.}, \au{Fukagata, K.} \& \au{Taira, K.}}
  \yr{2019{\natexlab{{\em a\/}}}}  \at{Super-resolution reconstruction of
  turbulent flows with machine learning}.  \jt{J. Fluid Mech.}  \bvol{870},
  \pg{106--120}.

\bibitem[Fukami {\em et~al.\/}(2020)Fukami, Fukagata \&
  Taira]{fukami2020machine}
{\sc \au{Fukami, K.}, \au{Fukagata, K.} \& \au{Taira, K.}} \yr{2020}
  \at{Machine learning based spatio-temporal super resolution reconstruction of
  turbulent flows}.  \jt{arXiv preprint arXiv:2004.11566} .

\bibitem[Fukami {\em et~al.\/}(2019{\natexlab{{\em b\/}}})Fukami, Nabae, Kawai
  \& Fukagata]{fukami_et_al}
{\sc \au{Fukami, Kai}, \au{Nabae, Yusuke}, \au{Kawai, Ken} \& \au{Fukagata,
  Koji}} \yr{2019{\natexlab{{\em b\/}}}}  \at{Synthetic turbulent inflow
  generator using machine learning}.  \jt{Phys. Rev. Fluids}  \bvol{4},
  \pg{064603}.

\bibitem[Fukushima(1980)]{fukushima1980neocognitron}
{\sc \au{Fukushima, K.}} \yr{1980}  \at{Neocognitron: A self-organizing neural
  network model for a mechanism of pattern recognition unaffected by shift in
  position}.  \jt{Biol. Cybern.}  \bvol{36}~(4),  \pg{193--202}.

\bibitem[Fukushima(1988)]{fukushima1988neocognitron}
{\sc \au{Fukushima, K.}} \yr{1988}  \at{Neocognitron: A hierarchical neural
  network capable of visual pattern recognition}.  \jt{Neural Networks}
  \bvol{1}~(2),  \pg{119--130}.

\bibitem[Guastoni {\em et~al.\/}(2019{\natexlab{{\em a\/}}})Guastoni, Encinar,
  Schlatter, Azizpour \& Vinuesa]{guastoni}
{\sc \au{Guastoni, L.}, \au{Encinar, M.~P.}, \au{Schlatter, P.}, \au{Azizpour,
  H.} \& \au{Vinuesa, R.}} \yr{2019{\natexlab{{\em a\/}}}}  \at{{Prediction of
  wall-bounded turbulence from wall quantities using convolutional neural
  networks}}.  \jt{J. Phys.: Conf. Ser.}  \bvol{1522},  \pg{012022}.

\bibitem[Guastoni {\em et~al.\/}(2019{\natexlab{{\em b\/}}})Guastoni,
  Srinivasan, Azizpour, Schlatter \& Vinuesa]{guastoni_tsfp}
{\sc \au{Guastoni, L.}, \au{Srinivasan, P.~A.}, \au{Azizpour, H.},
  \au{Schlatter, P.} \& \au{Vinuesa, R.}} \yr{2019{\natexlab{{\em b\/}}}} On
  the use of recurrent neural networks for predictions of turbulent flows.
  \bt{In {\em Proc. of the 11th Int. Symp. on Turbulence and Shear Flow
  Phenomena (TSFP11), Southampton, UK, July 30 - August 2\/}}.

\bibitem[G\"uemes {\em et~al.\/}(2019)G\"uemes, Discetti \&
  Ianiro]{Guemes2019sensing}
{\sc \au{G\"uemes, A.}, \au{Discetti, S.} \& \au{Ianiro, A.}} \yr{2019}
  \at{Sensing the turbulent large-scale motions with their wall signature}.
  \jt{Phys. Fluids}  \bvol{31}~(12),  \pg{125112}.

\bibitem[Ham {\em et~al.\/}(2019)Ham, Kim \& Luo]{ham_et_al_2019}
{\sc \au{Ham, Y.-G.}, \au{Kim, J.-H.} \& \au{Luo, J.~J.}} \yr{2019}  \at{{Deep
  learning for multi-year ENSO forecasts}}.  \jt{Nature}  \bvol{573},
  \pg{568--572}.

\bibitem[Hinton \& Salakhutdinov(2006)]{hinton2006reducing}
{\sc \au{Hinton, G.~E.} \& \au{Salakhutdinov, R.~R.}} \yr{2006}  \at{Reducing
  the dimensionality of data with neural networks}.  \jt{Science}
  \bvol{313}~(5786),  \pg{504--507}.

\bibitem[Hosseini {\em et~al.\/}(2016)Hosseini, Martinuzzi \&
  Noack]{hosseini_martinuzzi_noack_2016}
{\sc \au{Hosseini, Z.}, \au{Martinuzzi, R.~J.} \& \au{Noack, B.~R.}} \yr{2016}
  \at{Modal energy flow analysis of a highly modulated wake behind a
  wall-mounted pyramid}.  \jt{J. Fluid Mech.}  \bvol{798},  \pg{717--750}.

\bibitem[Ioffe \& Szegedy(2015)]{ioffe}
{\sc \au{Ioffe, S.} \& \au{Szegedy, C.}} \yr{2015} Batch normalization:
  Accelerating deep network training by reducing internal covariate shift.
  \bt{In {\em Proc. of the 32nd Int. Conf. on Machine Learning, 37\/}},
  \pg{pp. 448--456}.

\bibitem[Iten {\em et~al.\/}(2020)Iten, Metger, Wilming, Del~Rio \&
  Renner]{iten2020discovering}
{\sc \au{Iten, R.}, \au{Metger, T.}, \au{Wilming, H.}, \au{Del~Rio, L.} \&
  \au{Renner, R.}} \yr{2020}  \at{Discovering physical concepts with neural
  networks}.  \jt{Phys. Rev. Lett.}  \bvol{124},  \pg{010508}.

\bibitem[Jagodinski {\em et~al.\/}(2020)Jagodinski, Zhu \&
  Verma]{jagodinski2020uncovering}
{\sc \au{Jagodinski, E.}, \au{Zhu, X.} \& \au{Verma, S.}} \yr{2020}
  \at{Uncovering dynamically critical regions in near-wall turbulence using
  {3D} convolutional neural networks}.  \jt{arXiv preprint arXiv:2004.06187} .

\bibitem[Jean {\em et~al.\/}(2016)Jean, Burke, Xie, Davis, Lobell \&
  Ermon]{jean_et_al_2016}
{\sc \au{Jean, N.}, \au{Burke, M.}, \au{Xie, M.}, \au{Davis, W.~M.},
  \au{Lobell, D.~B.} \& \au{Ermon, S.}} \yr{2016}  \at{{Combining satellite
  imagery and machine learning to predict poverty}}.  \jt{Science}  \bvol{353},
   \pg{790--794}.

\bibitem[Jim{\'e}nez(2018)]{jimenez2018machine}
{\sc \au{Jim{\'e}nez, J.}} \yr{2018}  \at{Machine-aided turbulence theory}.
  \jt{J. Fluid Mech.}  \bvol{854}.

\bibitem[Kim \& Lee(2020)]{kim2020prediction}
{\sc \au{Kim, J.} \& \au{Lee, C.}} \yr{2020}  \at{Prediction of turbulent heat
  transfer using convolutional neural networks}.  \jt{J. Fluid Mech.}
  \bvol{882}.

\bibitem[Kingma \& Ba(2015)]{kingmaba}
{\sc \au{Kingma, D.~P.} \& \au{Ba, J}} \yr{2015} Adam: {A} method for
  stochastic optimization.  \bt{In {\em 3rd Int. Conf. on Learning
  Representations, {ICLR} 2015, San Diego, CA, USA, May 7-9, 2015\/}}.

\bibitem[Kutz(2017)]{kutz2017deep}
{\sc \au{Kutz, J.~N.}} \yr{2017}  \at{Deep learning in fluid dynamics}.  \jt{J.
  Fluid Mech.}  \bvol{814},  \pg{1--4}.

\bibitem[Lapeyre {\em et~al.\/}(2019)Lapeyre, Misdariis, Cazard, Veynante \&
  Poinsot]{lapeyre_et_al}
{\sc \au{Lapeyre, C.~J.}, \au{Misdariis, A.}, \au{Cazard, N.}, \au{Veynante,
  D.} \& \au{Poinsot, T.}} \yr{2019}  \at{Training convolutional neural
  networks to estimate turbulent sub-grid scale reaction rates}.  \jt{Combust.
  Flame}  \bvol{203},  \pg{255}.

\bibitem[LeCun {\em et~al.\/}(2015)LeCun, Bengio \& Hinton]{lecun2015deep}
{\sc \au{LeCun, Y.}, \au{Bengio, Y.} \& \au{Hinton, G.}} \yr{2015}  \at{Deep
  learning}.  \jt{Nature}  \bvol{521}~(7553),  \pg{436--444}.

\bibitem[LeCun {\em et~al.\/}(1989)LeCun, Boser, Denker, Henderson, Howard,
  Hubbard \& Jackel]{lecun1989backpropagation}
{\sc \au{LeCun, Y.}, \au{Boser, B.}, \au{Denker, J.~S.}, \au{Henderson, D.},
  \au{Howard, R.~E.}, \au{Hubbard, W.} \& \au{Jackel, L.~D.}} \yr{1989}
  \at{Backpropagation applied to handwritten {ZIP} code recognition}.
  \jt{Neural Comput.}  \bvol{1}~(4),  \pg{541--551}.

\bibitem[Lecun {\em et~al.\/}(1998)Lecun, Bottou, Bengio \&
  Haffner]{Lecun98gradient-basedlearning}
{\sc \au{Lecun, Y}, \au{Bottou, L.}, \au{Bengio, Y.} \& \au{Haffner, P.}}
  \yr{1998} Gradient-based learning applied to document recognition.  \bt{In
  {\em Proc. of the IEEE\/}},  \pg{pp. 2278--2324}.

\bibitem[Lee {\em et~al.\/}(1997)Lee, Kim, Babcock \& Goodman]{lee_et_al_cnn}
{\sc \au{Lee, C.}, \au{Kim, J.}, \au{Babcock, D.} \& \au{Goodman, R.}}
  \yr{1997}  \at{Application of neural networks to turbulence control for drag
  reduction}.  \jt{Phys. Fluids}  \bvol{9},  \pg{1740--1747}.

\bibitem[Ling {\em et~al.\/}(2016)Ling, Kurzawski \&
  Templeton]{ling2016reynolds}
{\sc \au{Ling, J.}, \au{Kurzawski, A.} \& \au{Templeton, J.}} \yr{2016}
  \at{Reynolds averaged turbulence modelling using deep neural networks with
  embedded invariance}.  \jt{J. Fluid Mech.}  \bvol{807},  \pg{155--166}.

\bibitem[Liu {\em et~al.\/}(2001)Liu, Adrian \& Hanratty]{liu2001large}
{\sc \au{Liu, Z.}, \au{Adrian, R.~J.} \& \au{Hanratty, T.~J.}} \yr{2001}
  \at{Large-scale modes of turbulent channel flow: transport and structure}.
  \jt{J. Fluid Mech.}  \bvol{448},  \pg{53--80}.

\bibitem[Long {\em et~al.\/}(2015)Long, Shelhamer \& Darrell]{long2015fully}
{\sc \au{Long, J.}, \au{Shelhamer, E.} \& \au{Darrell, T.}} \yr{2015} Fully
  convolutional networks for semantic segmentation.  \bt{In {\em Proc. of the
  IEEE Conf. on computer vision and pattern recognition\/}},  \pg{pp.
  3431--3440}.

\bibitem[Lozano-Dur{\'a}n {\em et~al.\/}(2020)Lozano-Dur{\'a}n, Bae \&
  Encinar]{lozano2020causality}
{\sc \au{Lozano-Dur{\'a}n, A.}, \au{Bae, H.~J.} \& \au{Encinar, M.~P.}}
  \yr{2020}  \at{Causality of energy-containing eddies in wall turbulence}.
  \jt{J. Fluid Mech.}  \bvol{882}.

\bibitem[Lozano-Dur{\'a}n {\em et~al.\/}(2012)Lozano-Dur{\'a}n, Flores \&
  Jim{\'e}nez]{lozano_2012}
{\sc \au{Lozano-Dur{\'a}n, A.}, \au{Flores, O.} \& \au{Jim{\'e}nez, J.}}
  \yr{2012}  \at{The three-dimensional structure of momentum transfer in
  turbulent channels}.  \jt{J. Fluid Mech.}  \bvol{694},  \pg{100--130}.

\bibitem[Lumley(1967)]{lumley1967structure}
{\sc \au{Lumley, J.~L.}} \yr{1967}  \at{The structure of inhomogeneous
  turbulent flows}.  \jt{Atmospheric turbulence and radio wave propagation} .

\bibitem[Marusic \& Heuer(2007)]{marusic_heuer}
{\sc \au{Marusic, I.} \& \au{Heuer, W. D.~C.}} \yr{2007}  \at{Reynolds number
  invariance of the structure inclination angle in wall turbulence}.  \jt{Phys.
  Rev. Lett.}  \bvol{99},  \pg{114504}.

\bibitem[Maulik {\em et~al.\/}(2019)Maulik, San, Rasheed \&
  Vedula]{maulik2019subgrid}
{\sc \au{Maulik, R.}, \au{San, O.}, \au{Rasheed, A.} \& \au{Vedula, P.}}
  \yr{2019}  \at{Subgrid modelling for two-dimensional turbulence using neural
  networks}.  \jt{J. Fluid Mech.}  \bvol{858},  \pg{122--144}.

\bibitem[Milano \& Koumoutsakos(2002)]{milano2002neural}
{\sc \au{Milano, M.} \& \au{Koumoutsakos, P.}} \yr{2002}  \at{Neural network
  modeling for near wall turbulent flow}.  \jt{J. Comput. Phys.}
  \bvol{182}~(1),  \pg{1--26}.

\bibitem[Moehlis {\em et~al.\/}(2004)Moehlis, Faisst \&
  Eckhardt]{moehlis_et_al}
{\sc \au{Moehlis, J.}, \au{Faisst, H.} \& \au{Eckhardt, B.}} \yr{2004}  \at{A
  low-dimensional model for turbulent shear flows}.  \jt{New J. Phys.}
  \bvol{6},  \pg{56}.

\bibitem[Mokhasi {\em et~al.\/}(2009)Mokhasi, Rempfer \&
  Kandala]{mokhasi_et_al}
{\sc \au{Mokhasi, P.}, \au{Rempfer, D.} \& \au{Kandala, S.}} \yr{2009}
  \at{Predictive flow-field estimation}.  \jt{Physica D}  \bvol{238},
  \pg{290--308}.

\bibitem[Murata {\em et~al.\/}(2020)Murata, Fukami \&
  Fukagata]{murata2020nonlinear}
{\sc \au{Murata, T.}, \au{Fukami, K.} \& \au{Fukagata, K.}} \yr{2020}
  \at{Nonlinear mode decomposition with convolutional neural networks for fluid
  dynamics}.  \jt{J. Fluid Mech.}  \bvol{882},  \pg{A13}.

\bibitem[Nair \& Goza(2020)]{nair_goza}
{\sc \au{Nair, N.~J.} \& \au{Goza, A.}} \yr{2020}  \at{Leveraging reduced-order
  models for state estimation using deep learning}.  \jt{J. Fluid Mech.}
  \bvol{897},  \pg{R1}.

\bibitem[Nair \& Hinton(2010)]{nair2010rectified}
{\sc \au{Nair, V.} \& \au{Hinton, G.~E.}} \yr{2010} Rectified linear units
  improve restricted {B}oltzmann machines.  \bt{In {\em Proc. of the 27th Int.
  Conf. on Machine Learning (ICML-10)\/}},  \pg{pp. 807--814}.

\bibitem[Norouzzadeh {\em et~al.\/}(2018)Norouzzadeh, Nguyen, Kosmala, Swanson,
  Palmer, Packer \& Clune]{norouzzadeh_et_al_2018}
{\sc \au{Norouzzadeh, M.~S.}, \au{Nguyen, A.}, \au{Kosmala, M.}, \au{Swanson,
  A.}, \au{Palmer, M.~S.}, \au{Packer, C.} \& \au{Clune, J.}} \yr{2018}
  \at{{Automatically identifying, counting, and describing wild animals in
  camera-trap images with deep learning}}.  \jt{Proc. Natl Acad. Sci.}
  \bvol{115},  \pg{E5716--E5725}.

\bibitem[van~den Oord {\em et~al.\/}(2016)van~den Oord, Dieleman, Zen,
  Simonyan, Vinyals, Graves, Kalchbrenner, Senior \&
  Kavukcuoglu]{oord2016wavenet}
{\sc \au{van~den Oord, A.}, \au{Dieleman, S.}, \au{Zen, H.}, \au{Simonyan, K.},
  \au{Vinyals, O.}, \au{Graves, A.}, \au{Kalchbrenner, N.}, \au{Senior, A.} \&
  \au{Kavukcuoglu, K.}} \yr{2016}  \at{Wavenet: A generative model for raw
  audio}.  \jt{arXiv preprint arXiv:1609.03499} .

\bibitem[Pan \& Yang(2009)]{pan2009survey}
{\sc \au{Pan, S.~J.} \& \au{Yang, Q.}} \yr{2009}  \at{A survey on transfer
  learning}.  \jt{IEEE Transactions on knowledge and data engineering}
  \bvol{22}~(10),  \pg{1345--1359}.

\bibitem[Pandey {\em et~al.\/}(2020)Pandey, Schumacher \&
  Sreenivasan]{pandey_et_al}
{\sc \au{Pandey, S.}, \au{Schumacher, J.} \& \au{Sreenivasan, K.~R.}} \yr{2020}
   \at{A perspective on machine learning in turbulent flows}.  \jt{J. Turbul.}
  \pg{pp. 1--18}.

\bibitem[Rabault {\em et~al.\/}(2019)Rabault, Kuchta, Jensen, R{\'e}glade \&
  Cerardi]{rabault2019artificial}
{\sc \au{Rabault, J.}, \au{Kuchta, M.}, \au{Jensen, A.}, \au{R{\'e}glade, U.}
  \& \au{Cerardi, N.}} \yr{2019}  \at{Artificial neural networks trained
  through deep reinforcement learning discover control strategies for active
  flow control}.  \jt{J. Fluid Mech.}  \bvol{865},  \pg{281--302}.

\bibitem[Raibaudo {\em et~al.\/}(2020)Raibaudo, Zhong, Noack \&
  Martinuzzi]{raibaudo2020machine}
{\sc \au{Raibaudo, C.}, \au{Zhong, P.}, \au{Noack, B.~R.} \& \au{Martinuzzi,
  R.~J.}} \yr{2020}  \at{Machine learning strategies applied to the control of
  a fluidic pinball}.  \jt{Phys. Fluids}  \bvol{32}~(1),  \pg{015108}.

\bibitem[Raissi {\em et~al.\/}(2019)Raissi, Perdikaris \&
  Karniadakis]{raissi2019physics}
{\sc \au{Raissi, M.}, \au{Perdikaris, P.} \& \au{Karniadakis, G.~E.}} \yr{2019}
   \at{Physics-informed neural networks: A deep learning framework for solving
  forward and inverse problems involving nonlinear partial differential
  equations}.  \jt{J. Comput. Phys.}  \bvol{378},  \pg{686--707}.

\bibitem[Raissi {\em et~al.\/}(2020)Raissi, Yazdani \&
  Karniadakis]{raissi_et_al}
{\sc \au{Raissi, M.}, \au{Yazdani, A.} \& \au{Karniadakis, G.~E.}} \yr{2020}
  \at{{Hidden fluid mechanics: Learning velocity and pressure fields from flow
  visualizations}}.  \jt{Science}  \bvol{367},  \pg{1026--1030}.

\bibitem[Sasaki {\em et~al.\/}(2019)Sasaki, Vinuesa, Cavalieri, Schlatter \&
  Henningson]{sasaki_vinuesa_cavalieri_schlatter_henningson_2019}
{\sc \au{Sasaki, K.}, \au{Vinuesa, R.}, \au{Cavalieri, A. V.~G.},
  \au{Schlatter, P.} \& \au{Henningson, D.~S.}} \yr{2019}  \at{Transfer
  functions for flow predictions in wall-bounded turbulence}.  \jt{J. Fluid
  Mech.}  \bvol{864},  \pg{708–745}.

\bibitem[Sirovich(1987)]{sirovich1987turbulence}
{\sc \au{Sirovich, L.}} \yr{1987}  \at{Turbulence and the dynamics of coherent
  structures. {II}. symmetries and transformations}.  \jt{Quart. Appl. Math.}
  \bvol{45}~(3),  \pg{573--582}.

\bibitem[Srinivasan {\em et~al.\/}(2019)Srinivasan, Guastoni, Azizpour,
  Schlatter \& Vinuesa]{srinivasan2019predictions}
{\sc \au{Srinivasan, P.~A.}, \au{Guastoni, L.}, \au{Azizpour, H.},
  \au{Schlatter, P.} \& \au{Vinuesa, R.}} \yr{2019}  \at{Predictions of
  turbulent shear flows using deep neural networks}.  \jt{Phys. Rev. Fluids}
  \bvol{4}~(5),  \pg{054603}.

\bibitem[Suzuki \& Hasegawa(2017)]{suzuki_hasegawa_2017}
{\sc \au{Suzuki, T.} \& \au{Hasegawa, Y.}} \yr{2017}  \at{Estimation of
  turbulent channel flow at ${Re}_\theta=100$ based on the wall measurement
  using a simple sequential approach}.  \jt{J. Fluid Mech.}  \bvol{830},
  \pg{760–796}.

\bibitem[Tinney {\em et~al.\/}(2006)Tinney, Coiffet, Delville, Hall, Jordan \&
  Glauser]{tinney2006spectral}
{\sc \au{Tinney, C.~E.}, \au{Coiffet, F.}, \au{Delville, J.}, \au{Hall, A.~M.},
  \au{Jordan, P.} \& \au{Glauser, M.~N.}} \yr{2006}  \at{On spectral linear
  stochastic estimation}.  \jt{Exp. Fluids}  \bvol{41}~(5),  \pg{763--775}.

\bibitem[Tinney {\em et~al.\/}(2008)Tinney, Ukeiley \& Glauser]{tinney2008low}
{\sc \au{Tinney, C.~E.}, \au{Ukeiley, L.~S.} \& \au{Glauser, M.~N.}} \yr{2008}
  \at{Low-dimensional characteristics of a transonic jet. {P}art 2. {E}stimate
  and far-field prediction}.  \jt{J. Fluid Mech.}  \bvol{615},  \pg{53--92}.

\bibitem[Udrescu \& Tegmark(2020)]{udrescu}
{\sc \au{Udrescu, S.-M.} \& \au{Tegmark, M.}} \yr{2020}  \at{{AI Feynman: A
  physics-inspired method for symbolic regression}}.  \jt{Sci. Adv.}  \bvol{6},
   \pg{1--16}.

\bibitem[Vinuesa {\em et~al.\/}(2020)Vinuesa, Azizpour, Leite, Balaam, Dignum,
  Domisch, Fell{\"a}nder, Langhans, Tegmark \& Fuso~Nerini]{vinuesa_et_al_2020}
{\sc \au{Vinuesa, R.}, \au{Azizpour, H.}, \au{Leite, I.}, \au{Balaam, M.},
  \au{Dignum, V.}, \au{Domisch, S.}, \au{Fell{\"a}nder, A.}, \au{Langhans,
  S.~D.}, \au{Tegmark, M.} \& \au{Fuso~Nerini, F.}} \yr{2020}  \at{{The role of
  artificial intelligence in achieving the Sustainable Development Goals}}.
  \jt{Nat. Commun.}  \bvol{11},  \pg{233}.

\bibitem[Wang {\em et~al.\/}(2017)Wang, Wu \& Xiao]{wang2017physics}
{\sc \au{Wang, J.~X.}, \au{Wu, J.~L.} \& \au{Xiao, H.}} \yr{2017}
  \at{{Physics-informed machine learning approach for reconstructing Reynolds
  stress modeling discrepancies based on DNS data}}.  \jt{Phys. Rev. Fluids}
  \bvol{2}~(3),  \pg{034603}.

\bibitem[Wu {\em et~al.\/}(2018)Wu, Xiao \& Paterson]{wu_et_al}
{\sc \au{Wu, J.-L.}, \au{Xiao, H.} \& \au{Paterson, E.}} \yr{2018}
  \at{{Physics-informed machine learning approach for augmenting turbulence
  models: A comprehensive framework}}.  \jt{Phys. Rev. Fluids}  \bvol{3},
  \pg{074602}.

\bibitem[Wu(2014)]{wu2014study}
{\sc \au{Wu, Y.}} \yr{2014}  \at{A study of energetic large-scale structures in
  turbulent boundary layer}.  \jt{Phys. Fluids}  \bvol{26}~(4),  \pg{045113}.

\end{thebibliography}

\end{document}